\documentclass[%
 aip,
 amsmath,amssymb,
 reprint,%
]{revtex4-2} 

\usepackage{graphicx}
\usepackage{dcolumn}
\usepackage{bm}

\usepackage[utf8]{inputenc}
\usepackage[T1]{fontenc}
\usepackage{mathptmx}
\usepackage{etoolbox}

\makeatletter
\def\@email#1#2{%
 \endgroup
 \patchcmd{\titleblock@produce}
  {\frontmatter@RRAPformat}
  {\frontmatter@RRAPformat{\produce@RRAP{*#1\href{mailto:#2}{#2}}}\frontmatter@RRAPformat}
  {}{}
}%
\makeatother
\begin{document}

\preprint{AIP/Chaos}

\title[Stochastic approach for assessing the chaotic predictability]{Stochastic approach for assessing the predictability of chaotic time series using reservoir computing}
\author{I.~A. Khovanov}
\email{i.khovanov@warwick.ac.uk}
\affiliation{ 
School of Engineering, University of Warwick, Coventry, CV4 7AL, UK
}%

\date{\today}

\begin{abstract}
The applicability of machine learning for predicting chaotic dynamics relies heavily upon the data used in the training stage. Chaotic time series obtained by numerically solving ordinary differential equations embed a complicated noise of the applied numerical scheme. Such a dependence of the solution on the numeric scheme leads to an inadequate representation of the real chaotic system. A stochastic approach for generating training times series and characterising their predictability is suggested to address this problem. The approach is applied for analysing two chaotic systems with known properties,  Lorenz system and Anishchenko-Astakhov generator. Additionally, the approach is extended to critically assess a reservoir computing model used for chaotic time series prediction. Limitations of reservoir computing for surrogate modelling of chaotic systems are highlighted.		
\end{abstract}

\maketitle

\begin{quotation}
The progress in machine learning techniques and methods makes them a tempting alternative to traditional modelling based on expert knowledge. For example, the relative simplicity of the reservoir computing technique leads to a remarkable range of its applications, from natural voice generation to turbulence prediction. Such a broad range could make an impression that reservoir computing is universally applicable for representing the system's dynamics. Therefore, a critical assessment and benchmarking of the machine learning technique are essential. Due to their complexity, chaotic systems provide a challenging benchmark for the technique. Since reservoir computing is a data-driven method, the time series used in the training stage should reflect the properties of a natural chaotic system. This paper illustrates that a typical numerical chaotic solution of ordinary differential equations significantly depends on the applied numerical scheme. Whilst this dependence is well-known, it is rarely taking into account. Using such a numerical solution for developing a reservoir computing model implies that the model represents a mixture of the system's dynamics and noise of the numerical scheme. In such a case, the extrapolation of the model's outcome for predicting the dynamics of a real system is questionable. A stochastic approach to generate data for machine learning methods is suggested to eliminate the influence of numerical noise. The approach includes explicitly stochastic terms that represent the dynamics of a natural system adequately. Furthermore, this approach provides a robust boundary of the predictability horizon for a given chaotic system. Therefore, to develop data-driven models, stochastic time series must be used, and models’ predictability must be compared with that obtained via the stochastic approach. Additionally, it is essential to understand the chaotic system's properties that data-driven models accumulate to be considered a ``true'' model. The consideration of two systems with distinct chaos structures shows significant limitations of a reservoir computing model for replacing one based on expert knowledge.	
\end{quotation}

\section{\label{sec1} Introduction}

Industry 4.0 revolution involves a broad application of cyber-physical systems and embeds models in engineering design. The latter requires transferring the expert knowledge and experience accumulated by the engineering community in the corresponding models. The traditional design based on linear models and decomposition of the whole system into smaller hierarchical subsystems needs to be adjusted to nonlinear characteristics and subsystems interactions. Therefore developing a comprehensive engineering model is a challenging and time-consuming task. The task is particularly difficult when nonlinearity plays the dominant role, for example, in the design and applications of NEMS \cite{Lifshitz:08}. Also, nonlinear models are rare in the engineering curriculum, which slows down embedding nonlinearity in engineering design. In such a situation, the development of data-driven models using artificial intelligence and machine learning techniques looks like an attractive alternative. Significant resources pouring into developing autonomous vehicles is one of many striking examples. An additional factor for advancing machine learning models consists in searching for novel unconventional computational paradigms. One of such paradigms is reservoir computing (RC).

RC implements a random recursive network that mimics neuronal cell cultures\cite{Jaeger:04}. The key idea is that the network can be implemented in various hardware as an analogue neural circuit. Examples\cite{Tanaka:19} of the RC implementation include electronic circuits\cite{Appeltant:11,Du:17,Zhong:21}, spin-torque nano oscillators\cite{Romera:18}, photonic devices\cite{Larger:17, Antonik:17} and NEMS\cite{Dion:18}. Note that conventional computing is used in the majority of RC research. The relative simplicity of associated computations can explain the popularity of RC. A small set of output gain factors is adjusted using a computationally inexpensive and robust procedure. It was argued\cite{Jaeger:04} that a high-dimension of the random network could approximate any signal. The RC technique has demonstrated a remarkable performance for various signals, including chaotic and turbulent \cite{Jaeger:04,Pathak:18,Zimmermann:18}. Typically, a task of time series prediction is considered. This task has been extended to the surrogate modelling, where RC replaces the model for generating the system output\cite{Jaeger:04,Pathak:18,Zimmermann:18,Fan:20}. Recently, a more challenging task of generalising the machine learning models has been tackled\cite{Weng:19,Kong:21,Kobayashi:21}. Whilst RC showed great promise, RC outcome often lacks critical assessment.

Due to the complexity of chaotic systems, chaotic time series is often used to benchmark machine learning techniques. The techniques are complicated; their implementation is the main subject of consideration, and as a result, the properties of chaos get less attention. Typically \cite{Jaeger:04,Pathak:18,Zimmermann:18,Fan:20}, a chaotic time series is obtained by numerical solving the corresponding ordinary differential equations. However, how well this numerical time series represents a real chaotic system? This question is central to applying machine learning methods to practical problems but remains outside many considerations. Moreover, discussions of the chaotic systems' complexity are often reduced to the analysis of the system’s sensitivity to the initial conditions. Such a reduction undermines the benchmarking itself. Thus the question of what chaotic time series can represent the output of a real chaotic system remains open.  Another related question is which properties of dynamical chaos should the machine learning model represent. For example, the RC model is a nonlinear and high-dimensional dynamical system, so RC dynamics can be complex, and adjusting the output gain factors provides a time series representation of RC high-dimensional state space. Additionally, the RC model has several parameters for tuning during the learning (training) process and initialisation. As a result, infinitely many RC approximations represent the time series and the corresponding dynamical system. Such an approximation can not be an exact copy of the analysing system. Some properties are preserved, but some are lost. Note that the comprehensive description of a given chaotic system is a challenge that requires an extensive analysis of bifurcations that leads to chaotic behaviour. There are different types of chaotic attractors with distinct manifestation. Widely considered Lorenz attractor is a quiasi-hyperbolic one. The attractor is robust to parameters uncertainties and external perturbations (noise). The structure of Lorenz attractor is well-understood\cite{Jackson:90}. A typical chaotic system is dramatically different, and there is no consensus on what a chaotic attractor is. A notion of a quasi-attractor as a combination of stable and unstable sets is widely applied. The structure of the quasi-attractor undergoes an infinite number of bifurcations as the system’s parameters vary \cite{Gonchenko:93, Gonchenko:08}. A co-existence of several regular and chaotic sets are often observed, and a typical chaotic system is structurally unstable. The noise significantly changes the properties of the quasi-attractor. The noise influence is particularly important for practical consideration since noise is abundant in real systems. Note that RC implementations also include noise. So typical chaotic and RC systems are stochastic, and this should be taking into account.

This paper aims to consider some of the questions formulated above. More specifically, first, an RC model is specified. Second, the RC predictability of chaotic time series is critically revised using two chaotic systems. One system is Lorenz model \cite{Lorenz:63}. The second one is a generator with inertial nonlinearity (GIN) of Anishchenko–Astakhov \cite{Anishchenko:82,Anishchenko:83}.  GIN system is comprehensively described in Anishchenko's book ``Complex Oscillations in Simple Systems'' \cite{Anishchenko:90}. An approach for generating chaotic time series for the predictability assessment is suggested. The approach aims to overcome the pitfalls of the numerical methods for solving ordinary differential equations by adding explicit stochastic components into the system. Further, the surrogate modelling and the generalisation of a machine learning model are discussed for these two systems. Finally, the results are summarised and discussed in the conclusion section.

\section{\label{sec2}  Reservoir computing}

Among all possible implementations of RC, the echo state network (ESN) of leaky integrator neurons is the most popular\cite{Pathak:18,Zimmermann:18,Fan:20,Weng:19,Kong:21}. A dynamical reservoir of neurons is described by the following equation:
\begin{eqnarray}
\dot{{\mathbf v}} = \frac{1}{\tau} \left( - {\mathbf v} + \tanh ({\mathbf W}_i {\mathbf u} + {\mathbf W}{\mathbf v} )  \right)
\label{eq1}
\end{eqnarray}
where vector ${\mathbf v}=(v_1, v_2,\ldots, v_n)$ describes the state of $n$ neurons with same decay constant $\tau$, ${\mathbf u}=(u_1, u_2,\ldots, u_l)$ is neuron's input, matrices ${\mathbf W}_i$  and ${\mathbf W}$ define input and internal weights, respectively. An additional layer forms the recurrence in the network: 
\begin{eqnarray}
\dot {\mathbf u} = {\mathbf W_{ou}} {\mathbf u} + {\mathbf W_{ov}} {\mathbf v} 
	\label{eq2}
\end{eqnarray}
with matrices ${\mathbf W}_{ou}$  and ${\mathbf W_{ov}}$ defining output weights. In a conventional computer, the differential equations (\ref{eq1}) and (\ref{eq2}) are used in the form of the following discrete-time system (map):
\begin{eqnarray}
{\mathbf v}_{k+1} & = & (1-\alpha){\mathbf v}_k + \tanh ({\mathbf W}_i {\mathbf u}_k + {\mathbf W}{\mathbf v}_k ) \ ,	\label{eq3} \\
 {\mathbf u}_{k+1} & = & {\mathbf W}_o [ {\mathbf u}_k {\mathbf v}_k ] .
	\label{eq4}
\end{eqnarray}
This map is formally obtained by applying the explicit Euler method \cite{Butcher:16} with the step size $h$. Parameter $\alpha \in(0,1]$ is the leaking rate which links to the decay constant via the expression $\alpha=\frac{h}{\tau}$. It is known\cite{Butcher:16} that the explicit Euler method is conditionally stable, and the step size controls the stability. Therefore, adjusting parameter $\alpha$ affects the neuron's decay rate and the stability of maps (\ref{eq3}) and (\ref{eq4}) simultaneously. It means that the selection of the value of $h$ is irrelevant in the transition from continuous to discrete time. However, $h$ plays a crucial role in the opposite transition to an analogue realisation of the ESN.

Matrix ${\mathbf W}_i$ is random, having values within a particular range for mapping input ${\mathbf u}$ into the linear range of the hyperbolic tangent. Similar mapping is applied to matrix ${\mathbf W}$ with additional constraint on the spectral radius\cite{Jaeger:04}, $\rho({\mathbf W})\le 1$. Additionally, it is recommended\cite{Jaeger:04} to use a sparse matrix ${\mathbf W}$. The learning process consists of determining matrix  ${\mathbf W}_o$ via the least-square regression for a given input time series ${\mathbf s}$, used for substitution ${\mathbf u}={\mathbf s}$. The input time series is scaled. During learning, the leaking rate, $\alpha$, and spectral radius, $\rho({\mathbf W})$, are adjusting parameters for the network's stability and longer prediction. The successful application of RC requires experience and following some tips \cite{Lukosevicius:12}. Several codes templates are freely available, and the Matlab script from the website https://mantas.info/ was used in this paper.

\section{\label{sec3}  On the prediction of chaotic time series}

\subsection{\label{sec3.1}  Lorenz and GIN systems}

Lorenz and GIN systems are selected for analysis. Both are three dimensional and well studied. The differential equations for Lorenz system are
\begin{eqnarray}
	\dot{x} & = & p (y-x) \nonumber \\
    \dot{y} & = & -x z+rx-y  	\label{eq5}\\
    \dot{z} & = & xy-bz \nonumber
\end{eqnarray}
where $p$, $r$ and $b$ are parameters. The following parameters values are used: $p=10$, $r=28$ and $b=8/3$. GIN system has the following form:
\begin{eqnarray}
	\dot{x} & = & m x+y-x z \nonumber \\
	\dot{y} & = & -x  	\label{eq6}\\
	\dot{z} & = & -g z +g I(x) x^2 \nonumber  
\end{eqnarray}
where $m$ and $g$ are parameters; $m=2.412$ and $g=0.097$. Function $I(x)$ is defined by the following expression:
\begin{eqnarray}
	I(x)= \left \{ 
		\begin{array}{ll}
		1, & x\ge0 \\
		0, &x<0 
		\end{array}
		\right. \nonumber
\end{eqnarray}

\subsection{\label{sec3.2}  Pitfalls of numerical solving chaotic systems}

\begin{figure}
\noindent \includegraphics[width=1.6in]{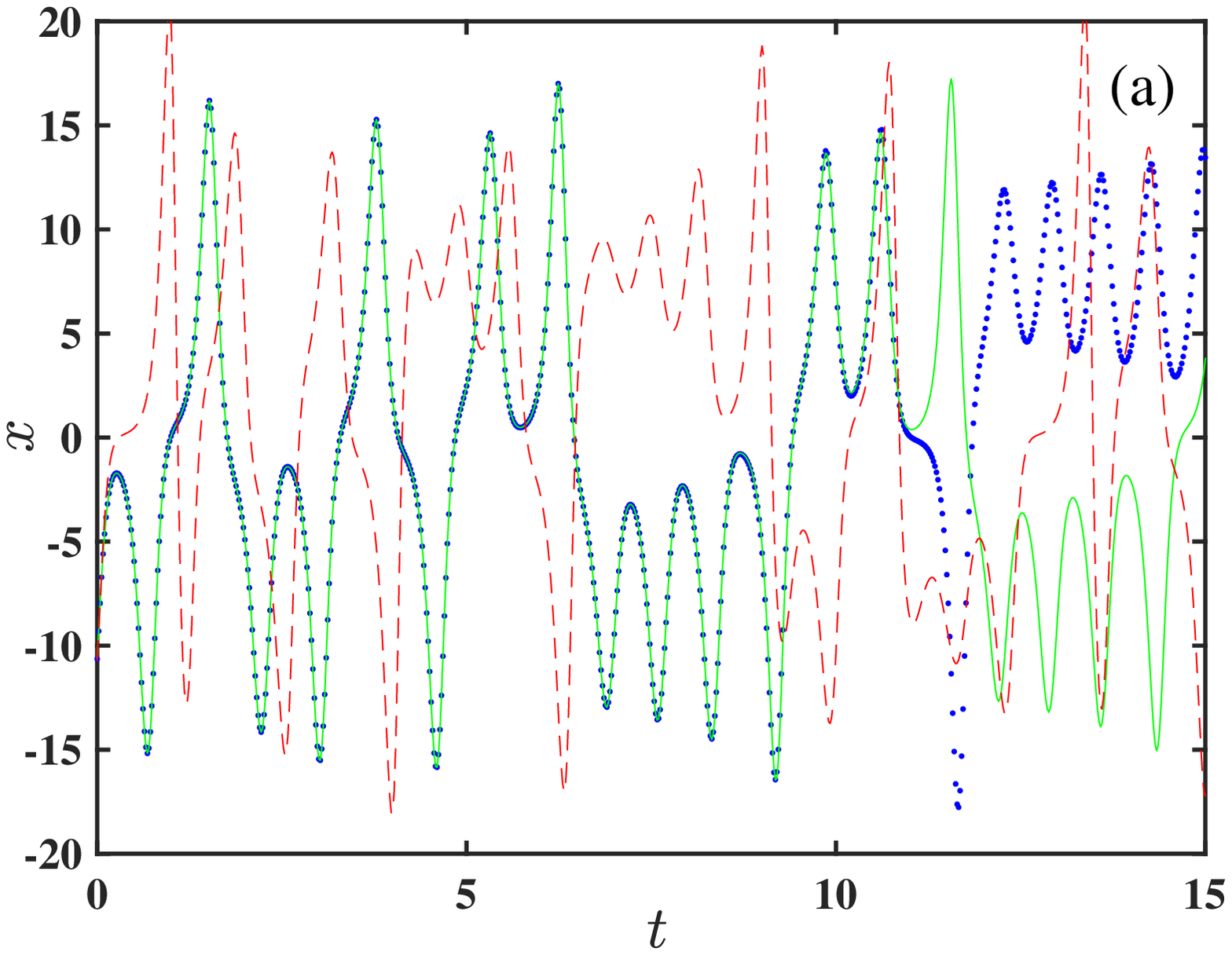}~~~\includegraphics[width=1.6in]{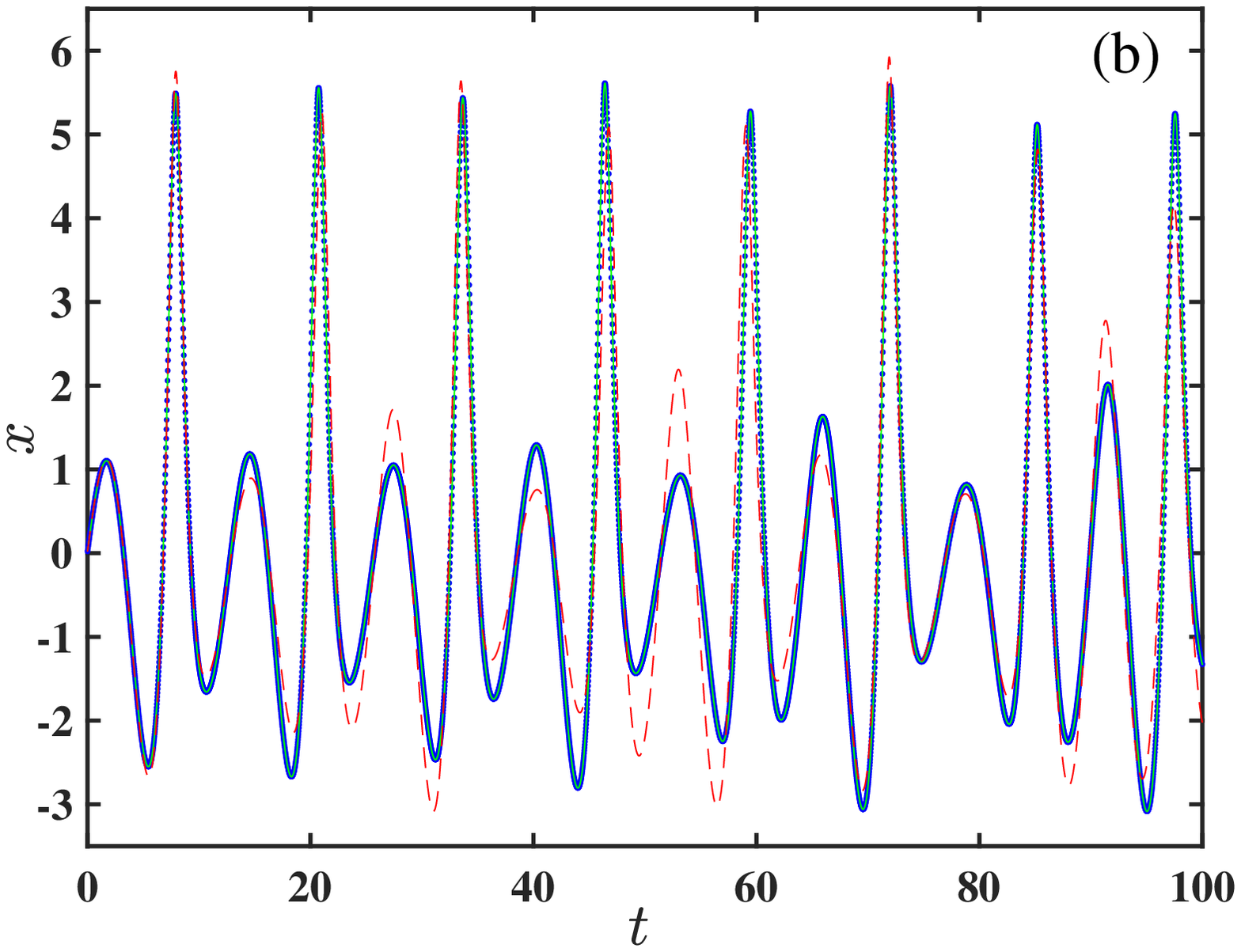}\\
\noindent \includegraphics[width=1.6in]{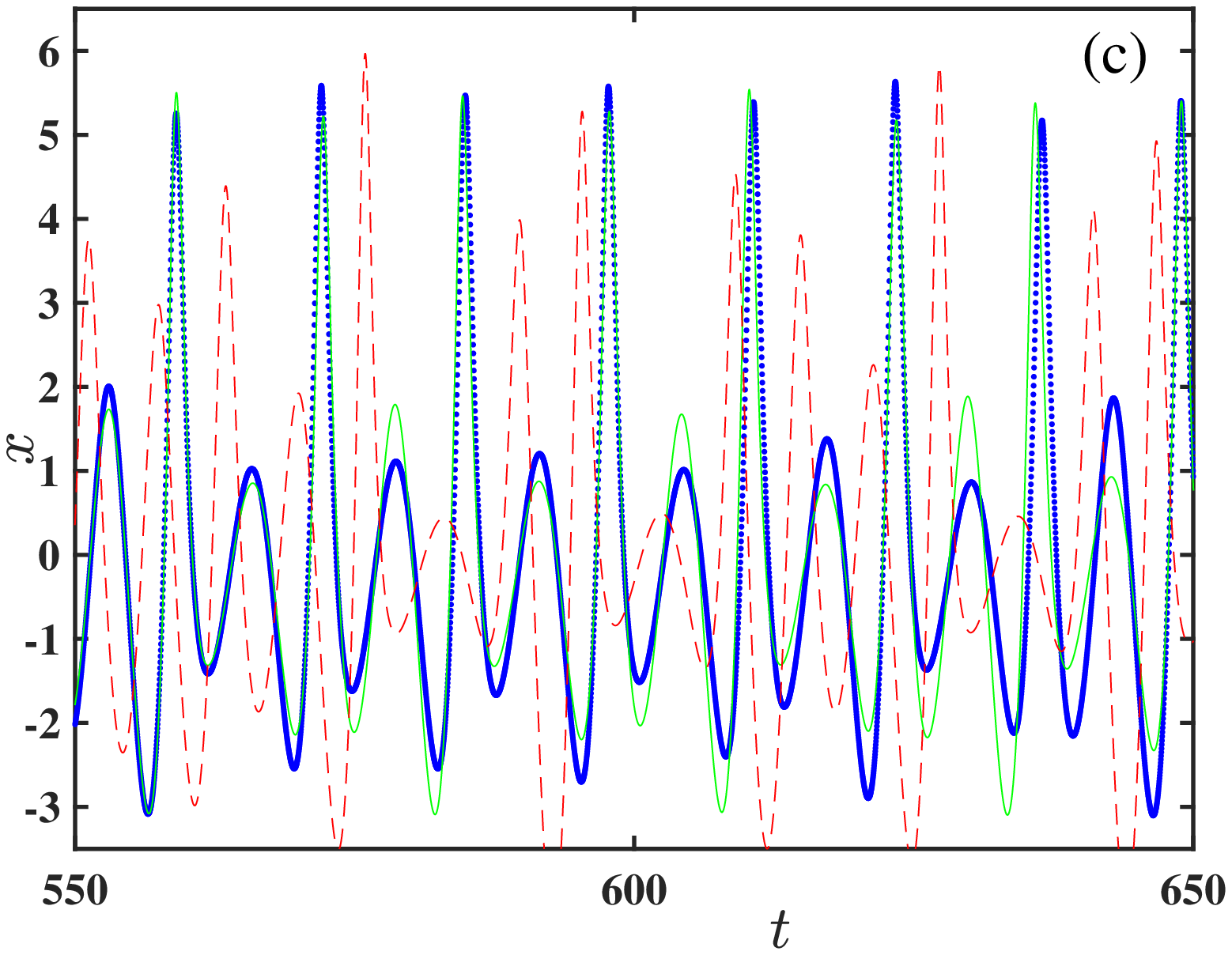}~~~\includegraphics[width=1.6in]{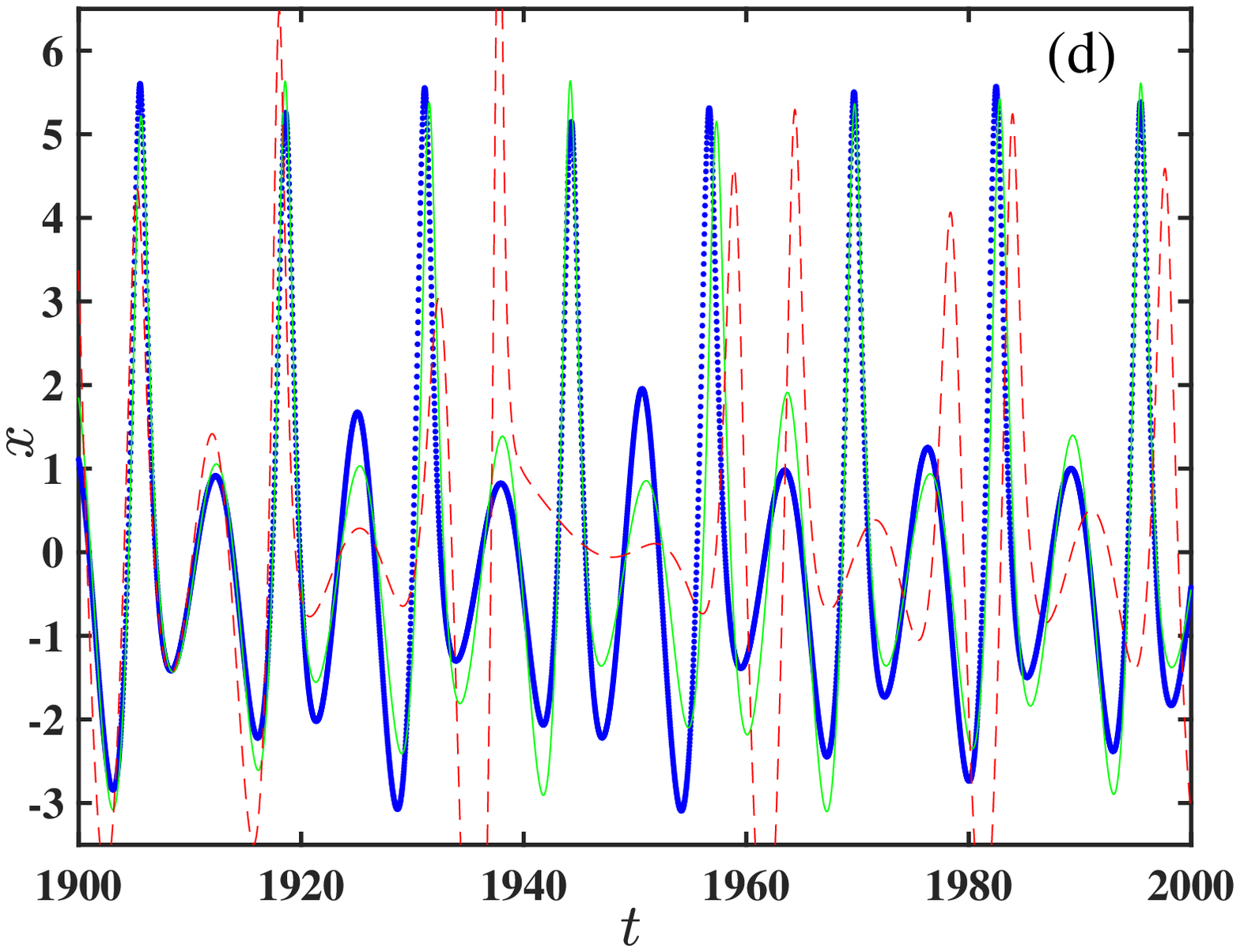}
	\caption{\label{fig1} Numerical solutions obtained by Euler method (red dashed line), the classical 4th order Runge-Kutta method (blue dotted line) and Ralston method (green solid  line) are shown for Lorenz system (a) and GIN system (b, c and d). The step size, $h=0.02$, was used for all numerical methods. }
\end{figure}

In the majority of publications considering RC, chaotic time series were obtained by numerical solving the corresponding ordinary differential equations. Standard numerical methods implemented in software packages or custom codes were used. Typically, these methods are based on the Runge-Kutta approach \cite{Butcher:16}. Solutions of systems (\ref{eq5}) and (\ref{eq6}) obtained using three different Runge-Kutta methods with the same integration step size $h=0.02$ are shown in Fig.~\ref{fig1}. These time series confirm that chaotic systems demonstrate sensitivity to numerical methods. It is apparent that a given numerical chaotic time series is an approximation of an unknown ``true'' solution. The approximation's properties depend significantly on the applied numerical methods. More strictly, a numerical solution of chaotic systems does not guarantee that the numerical approximation is close to the true solution. Therefore the analysis of a chaotic system must include an additional consideration using bifurcations of simpler sets, for example, equilibrium points, cycles and their manifolds\cite{Arnold:94}, or approximate analytical methods \cite{Soskin:08,Soskin:15}.

The differences in the applied numerical schemes lead to the mismatch between time series (Fig.~\ref{fig1}). One of the schemes is the explicit Euler method, which has a low-order accuracy with the global error of $O(h)$. Two other schemes have the same fourth-order accuracy $O(h^4)$ but different coefficients. One of them is the classical 4th order Runge-Kutta scheme\cite{Kutta:01}, and another is known as the Ralston scheme aiming to minimize the truncation error\cite{Ralston:62}. Runge-Kutta methods are based on Taylor series with the corresponding order truncation. The truncation leads to an error at every integration step. This integration error depends on the system's state $(x,y,z)$. The round-off errors of computations also contribute to the numerical error. As a result, numerical solving of a deterministic system (ordinary differential equations) leads to a stochastic map with multiplicative (state dependent) noise with complicated properties. The numerical noise could lead to different effects depending on the structure of a chaotic set. In Fig.~\ref{fig1}(a), parameters of Lorenz system are selected in the range of a quasi-hyperbolic chaotic attractor with a well-defined structure. The noise influence for such attractors is expected to be minimal because of the shadowing properties of a hyperbolic chaotic set\cite{Sinai:72}. In GIN system, identifying the corresponding attracting chaotic set(s) is an open question. It is known\cite{Anishchenko:90} that this system has a quasi-attractor: a mixture of different stable and unstable sets. So, the numerical noise induces a diffusion between sets and, in this sense, this noise is an embedded feature of the observed chaotic motion. Note that GIN system is generic.

The Euler method has the lowest accuracy, and the resulting trajectories (red dashed lines in Fig.~\ref{fig1}) significantly differ from those obtained by high order schemes. The influence of the numerical noise is strong in the Euler method. For Lorenz system (Fig.~\ref{fig1}(a)), trajectories obtained by 4th order methods are close to each other until a particular time moment ($t\approx 11$). Then the trajectories deviate abruptly. This deviation happens in the region of splitting stable manifold of the saddle point of system (\ref{eq1}) \cite{Jackson:90}. It has been  shown\cite{Anishchenko:01,Anishchenko:02} that this region is responsible for large deviations from the chaotic attractor. Fig.~\ref{fig1}(a) shows that this region also amplifies the numerical noise. A different picture is observed for GIN system (Fig.~\ref{fig1}(b-c)). Trajectories of 4th order schemes are close to each other for long interval, and when the trajectories deviate from each other, they still stay close. The difference between trajectories oscillates and remains smaller than the range of coordinates $x$ (Fig.~\ref{fig1}(c)) for a long time interval. The presence of the state space regions with strong and weak dissipations in  GIN system\cite{Anishchenko:90} could explain the oscillations. Thus, in contrast to  Lorenz system, the numerical noise does not dramatically differentiate trajectories in GIN system. Nevertheless, the presence of numerical errors is apparent.

\subsection{\label{sec3.3}  Stochastic approach}

The pitfalls of the numerical solutions described earlier are vital for rigorous consideration of deterministic chaotic systems. Real systems include stochastic components, and their statistical properties are different from those of numerical noise. The pitfalls can be addressed by the explicit inclusion of stochastic additive terms into differential equations. I learnt this idea from Prof. Vadim Anishchenko a long time ago, but this idea has not been explored. So, a model in the form of stochastic differential equations (SDEs) is more realistic. If the noise intensity is larger than the level of the numerical noise, the latter becomes irrelevant. More importantly, the stochastic model stresses an approximate, random character of a single numerical solution. Also, the SDE model includes the step size value into consideration. Note that the selection of $h$ value is rarely discussed and often set arbitrary, especially among the machine learning community. Additionally, there is an opportunity to select the stochastic term in a particular form for reflecting the properties of a given real system. White Gaussian noise is a reasonable representation of stochasticity in many situations. 

The SDEs for Lorenz system are the following:
\begin{eqnarray}
	\dot{x} & = & p (y-x) +\sqrt{D_x} \xi_x(t) \nonumber \\
	\dot{y} & = & -x z+rx-y +\sqrt{D_y} \xi_y(t)  	\label{eq7}\\
	\dot{z} & = & xy-bz +\sqrt{D_z} \xi_z(t) \nonumber
\end{eqnarray}
and for GIN system, the SDEs are the following:
\begin{eqnarray}
	\dot{x} & = & m x+y-x z +\sqrt{D_x} \xi_x(t) \nonumber \\
	\dot{y} & = & -x  + \sqrt{D_y} \xi_y(t)	\label{eq8}\\
	\dot{z} & = & -g z +g I(x) x^2 + \sqrt{D_z} \xi_z(t) \nonumber  
\end{eqnarray}
where $D_x$, $D_y$ and $D_z$ are intensities of white Gaussian noises $\xi_x(t)$, $\xi_y(t)$ and $\xi_z(t)$, respectively. 

For particular initial conditions, a stochastic system shows infinitely many realisations. Therefore, a statistical analysis of an ensemble of system's trajectories (time series) must be applied. This analysis leads to the robust boundary for time series predictability. Additionally, a comparison between a deterministic system and its stochastic counterpart clarifies the role of numerical noise for the given system. Note that many numerical schemes can be extended to include stochastic terms.

\begin{figure}
\noindent \includegraphics[width=1.6in]{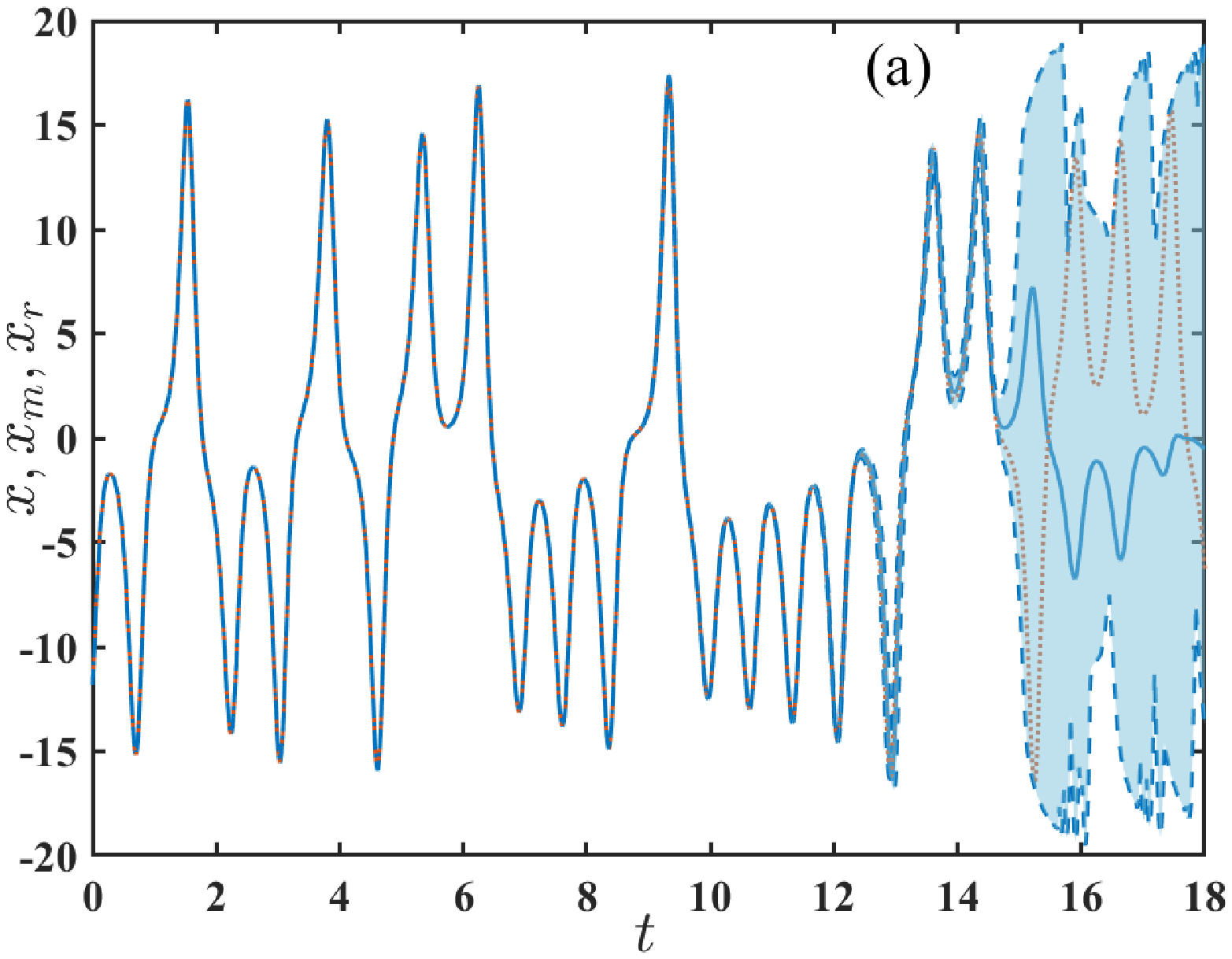}~~~\includegraphics[width=1.6in]{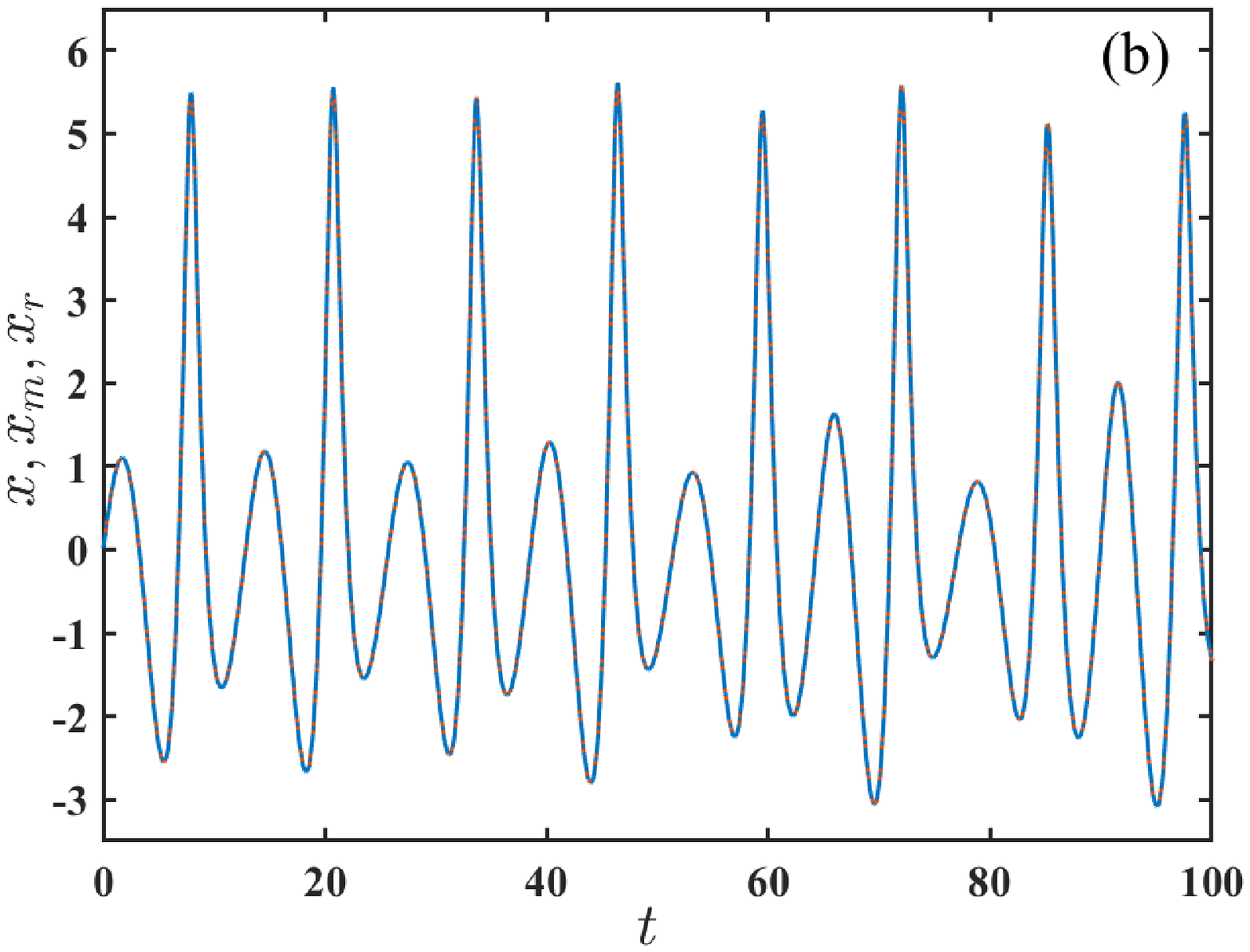}\\
\noindent \includegraphics[width=1.6in]{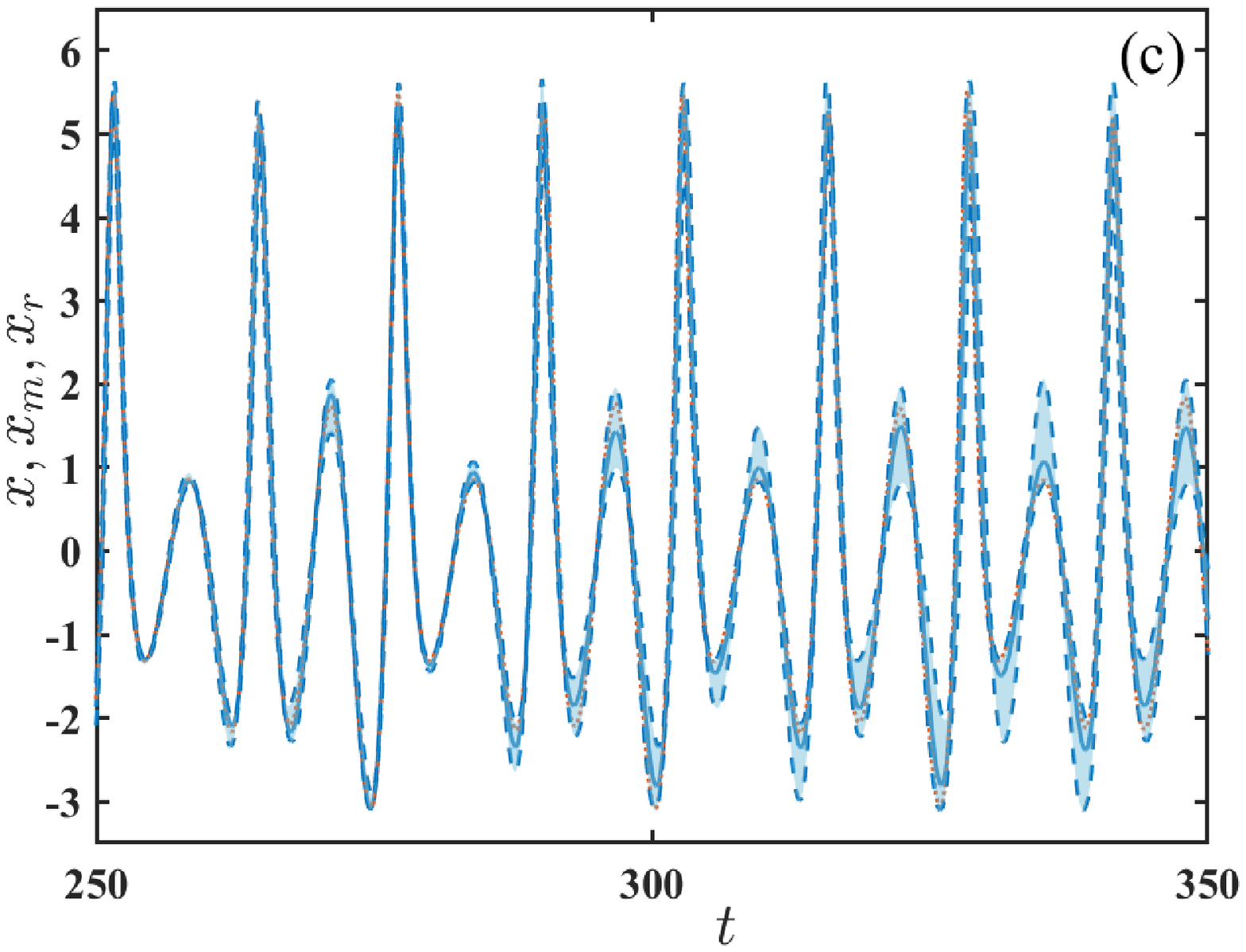}~~~\includegraphics[width=1.6in]{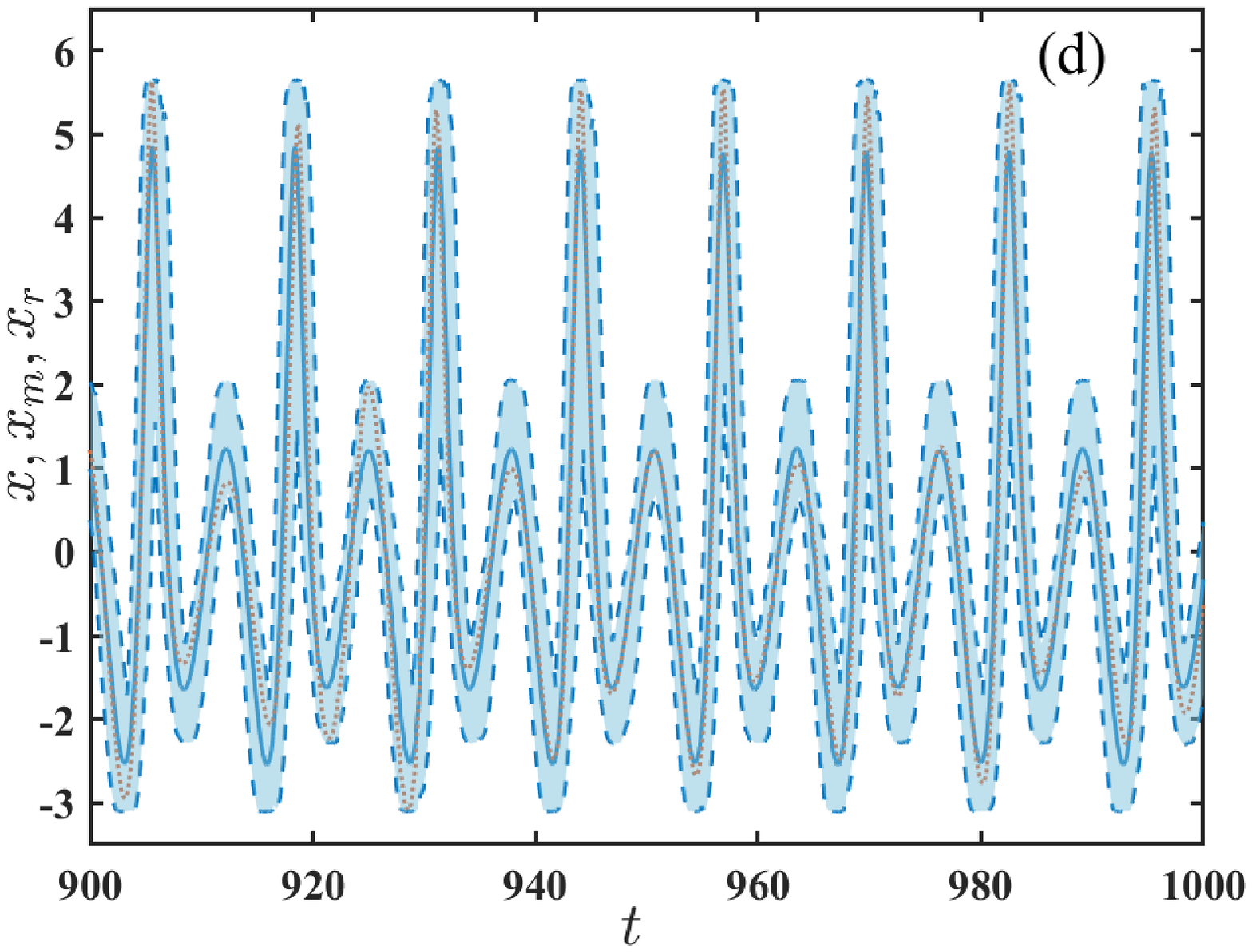}
	\caption{\label{fig2}  The application of the stochastic approach for characterising the predictability of Lorenz system (a) and GIN system (b, c and d) is illustrated. Time-dependent mean $x_m(t)$, and  range, $x_r(t)$, for the trajectories ensemble are shown by solid and dashed blue lines, respectively. The range, $x_r(t)$, is additionally highlighted by shadowed area. One of the ensemble trajectories is shown by red dotted line. The step size for SDEs is $h=0.001$ and noise intensities are $D_x=D_y=D_z=10^{-6}$. }
\end{figure}

Deterministic trajectories shown in Fig.~\ref{fig1} were obtained for the step size $h=0.02$. For stochastic systems (\ref{eq7}) and (\ref{eq8}) the step size should be smaller for eliminating the numerical noise, so the selected step size is $h_s=0.001$. Noise intensities should be larger than the magnitude of the numerical noise, which for the classical 4th order Runge Kutta scheme is $O(h^4)$. Therefore, the selected noise intensities are $D_x=D_y=D_z=10^{-6}$. Ensemble of 1000 trajectories started with the same initial conditions as those in Fig.~\ref{fig1} was used for the statistical description. The time evolution of the ensemble is characterised by the mean value and range of coordinate $x$. Note that in the RC prediction, a single coordinate (one state variable) is typically considered. So, for both systems (\ref{eq7}) and (\ref{eq8}), time-dependent mean, $x_m(t)$, and the range, $x_r(t)$, for the trajectories ensemble are calculated for coordinate $x(t)$. Figure \ref{fig2} illustrates a time evolution of the ensemble. Similar to the numerical noise in Fig.~\ref{fig1}, the stochastic perturbations lead to an abrupt deviation of trajectories in Lorenz system (Fig,~\ref{fig2}(a)) and oscillatory trajectories' difference in  GIN system (Fig.~\ref{fig2}(b-c)). However, the stochastic approach provides a quantitative description. The approach gives a robust estimation of the trajectories' predictability for given initial conditions.

\subsection{\label{sec3.4}  Predictability of chaotic time series}

\begin{figure}
\noindent \includegraphics[width=1.6in]{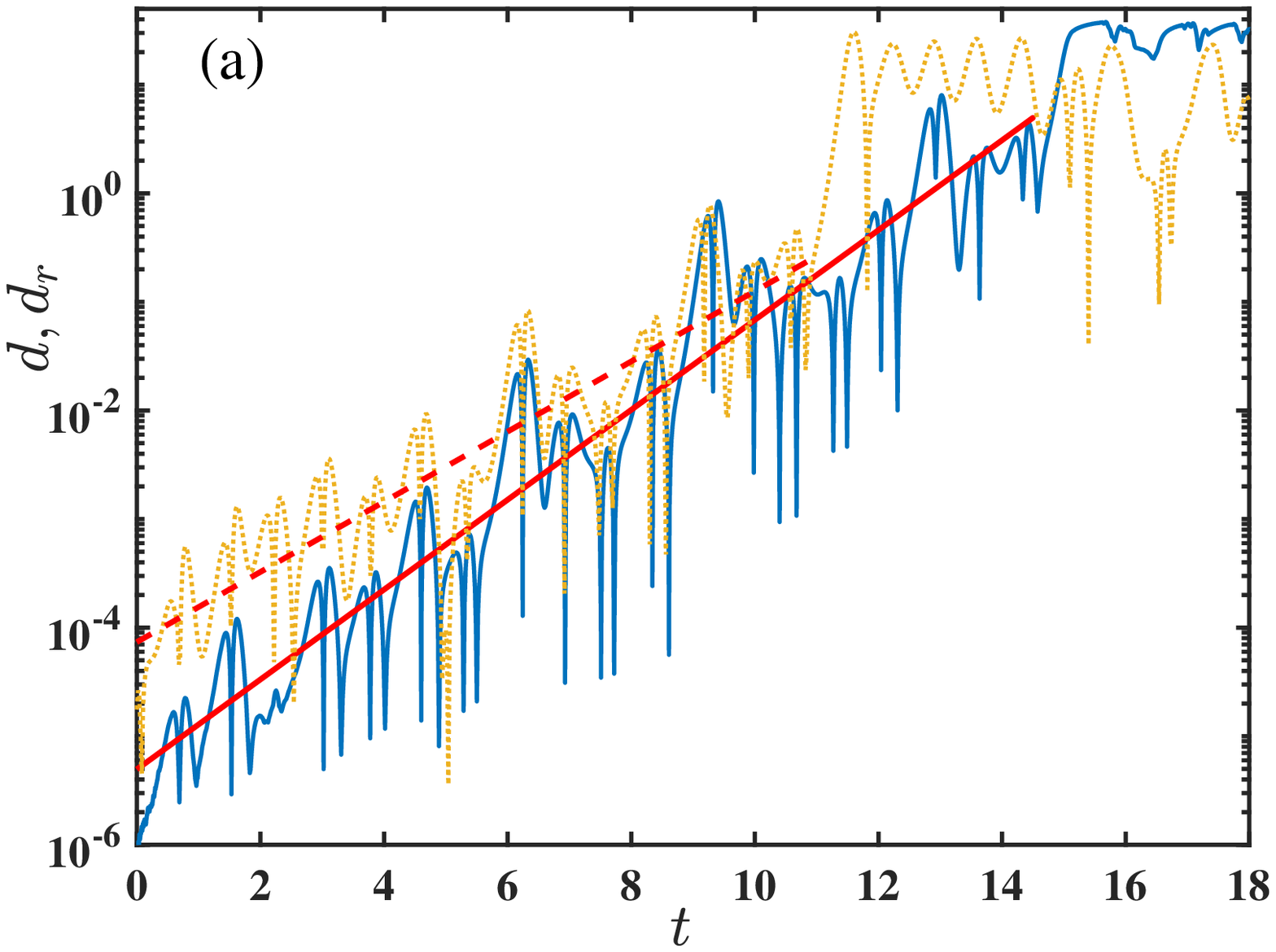}~~~\includegraphics[width=1.6in]{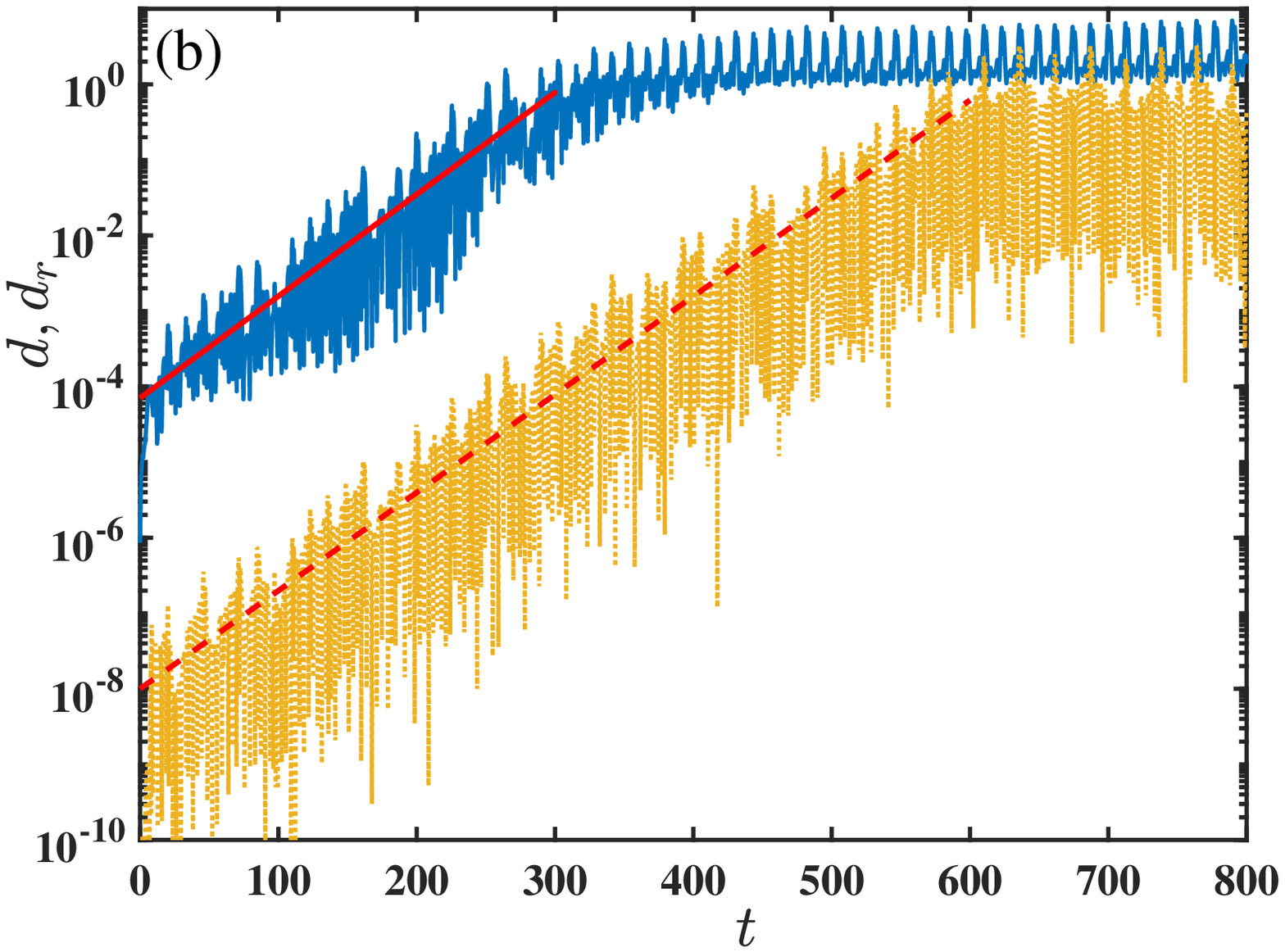}
	\caption{\label{fig3} The predictions errors $d(t)$ (orange dotted line) and $d_r(t)$ (blue solid line) are shown for Lorenz (a) and GIN (b) systems. The ordinate axis is shown in the logarithmic scale. Exponential scaling of both errors is illustrated with a fitting line.}
\end{figure}

In this section, the prediction horizons defined by the numerical noise in systems  (\ref{eq5}) and (\ref{eq6}) and by white noise in systems  (\ref{eq7}) and (\ref{eq8}) are compared. Two trajectories started for the same initial conditions but estimated using the classical Runge-Kutta and Ralston schemes are considered for systems (\ref{eq5}) and (\ref{eq6}). The time-dependent absolute error, $d(t)$, for coordinate, $x$, is calculated: $d(t)=| x_{RK}(t)-x_{R}(t)|$; subscripts $RK$ and $R$ correspond to the classical Runge-Kutta and Ralston schemes, respectively. The size, $d_r(t)$, of the range, $x_r(t)$, of trajectories ensemble is used for SDEs (\ref{eq7}) and (\ref{eq8}). The results are summarised in Fig.~\ref{fig3}.

For Lorenz system, both indicators $d(t)$ and $d_r(t)$ (Fig.\ref{fig3}(a)) show an initial exponential growth followed by a sharp increase at some time moments. After the increase, both indicators tend to a saturation value comparable with the range of coordinate $x$. Time moments, $\tau$, corresponding to the sharp increase of the indicators, are the prediction horizons. The value of $\tau$ was calculated using the conditions: $d(\tau)>0.1 x_m$ and $d_r(\tau)>0.1 x_m$; here $x_m$ is the maximal value of coordinate $x(t)$. For selected initial conditions, the horizon defined by the numerical noise is $\tau_n \approx 11$, and the horizon obtained using the stochastic framework is longer, $\tau_{sf} \approx 14$. Both indicators undergo a sharp increase when trajectories approached a region of splitting of stable manifold sheets. Note that the manifold structure is comprehensively described by Jackson\cite{Jackson:90}. The exponential growth interval can be used for estimating the largest Lyapunov exponent, $\lambda$. Indicator $d(t)$ gives value $\lambda \approx 0.744$, and the stochastic approach leads to value $ \lambda \approx 0.953$. The latter is close to the largest Lyapunov exponent, which is around $0.9$.

The time evolution of indicators $d(t)$ and $d_r(t)$ for GIN system is shown in Fig.~\ref{fig3}(b). An initial exponential growth leads to a saturation level around which both indicators oscillate. The stochastic approach shows that   trajectories do not spread over the whole chaotic attractor for a long period of time. There is an alternating pattern of spreading and contracting of trajectories. This pattern reflects the state space structure in GIN system having regions with instability and strong dissipation\cite{Anishchenko:90}. The prediction horizon, $\tau$, is defined by the same conditions as for  Lorenz system. The prediction horizons for the numerical and stochastic noises are $\tau_n \approx 600$ and $\tau_{sf} \approx 300$, respectively. However, the prediction error defined by $d(t)$ or $d_r(t)$ remains finite, and trajectories are close to each other for significantly longer than $\tau_n$ and $\tau_{sf}$ time intervals. For the stochastic ensemble, if the prediction horizon is defined as a time moment, $\tau_m$, when the minimal range ${\rm min} [ d_r(t) ] > 0.1 x_m$ for $t\ge \tau_m$, then $\tau_m\approx 3000$.
Note that the initial exponential growth of  $d(t)$ or $d_r(t)$ gives close estimations of the largest Lyapunov exponents: $\lambda=0.031$ and $\lambda=0.030$ for the stochastic and numerical noise, respectively. The largest Lyapunov exponent for system (\ref{eq6}) is approximately $0.027$. The results shown in Fig.~\ref{fig2} and Fig.~\ref{fig3}, confirm that the stochastic approach provides a robust estimation of the prediction horizon for both systems.

\begin{figure}
	\noindent \includegraphics[width=1.6in]{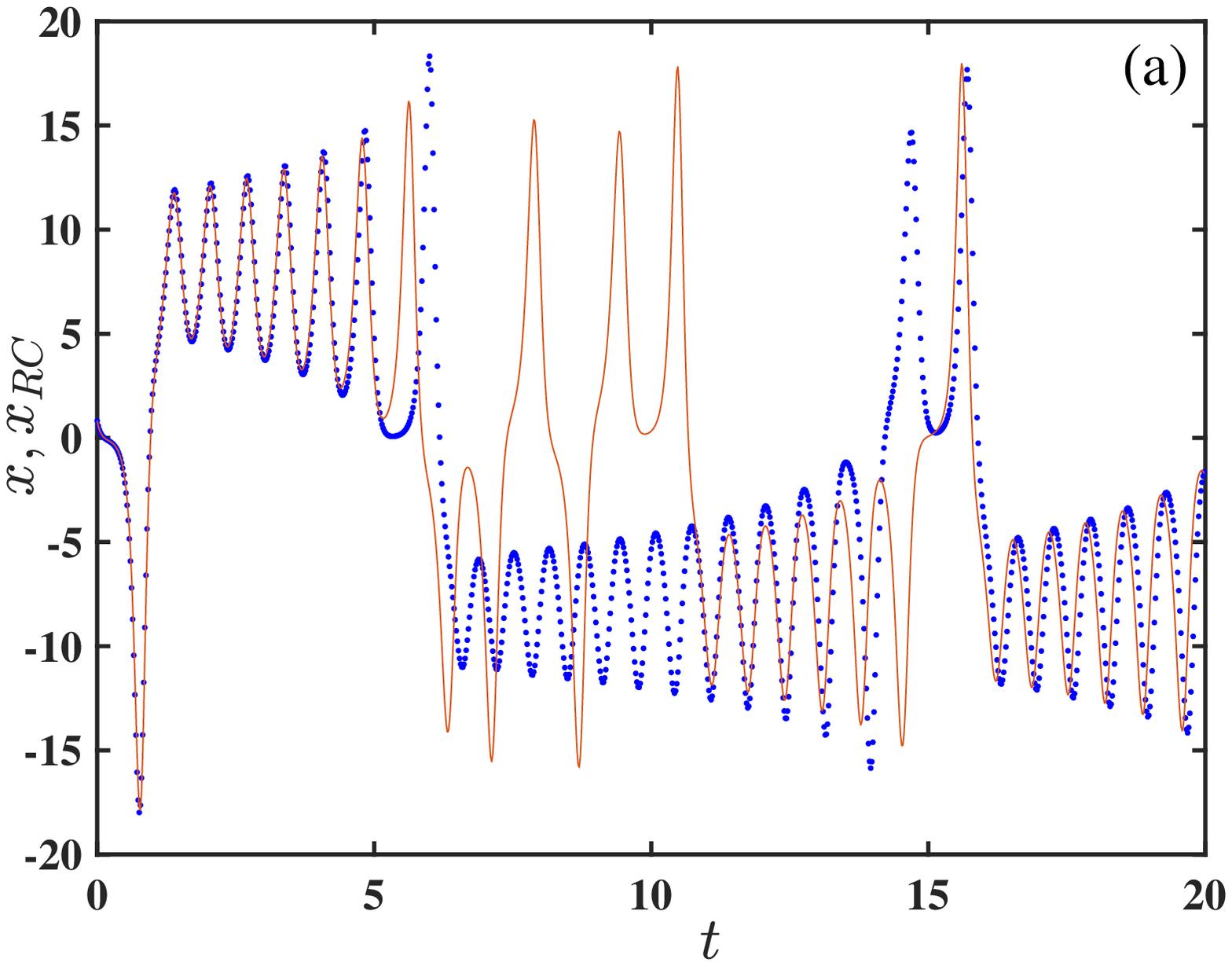}~\includegraphics[width=1.6in]{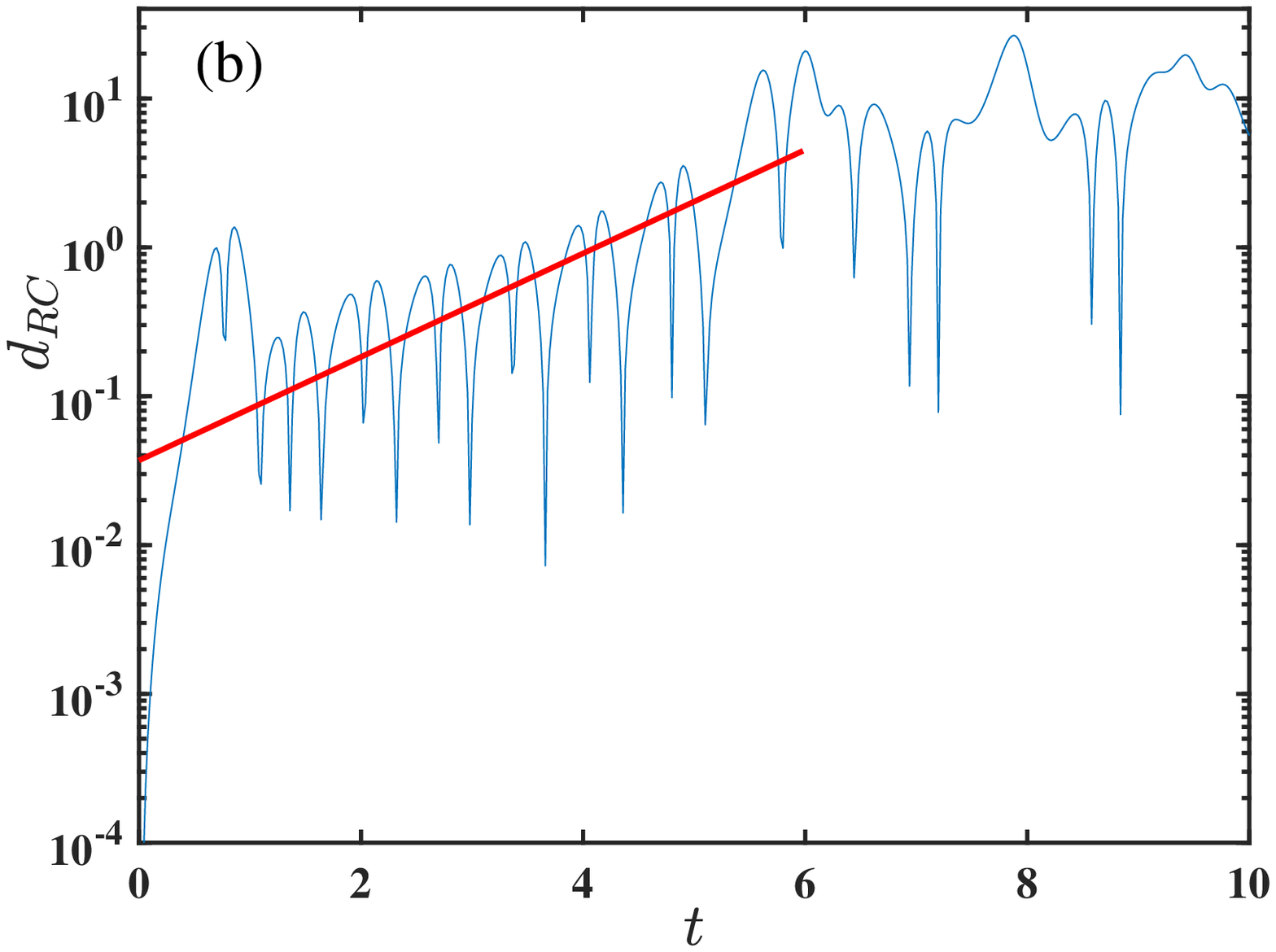}\\
	\noindent \includegraphics[width=1.6in]{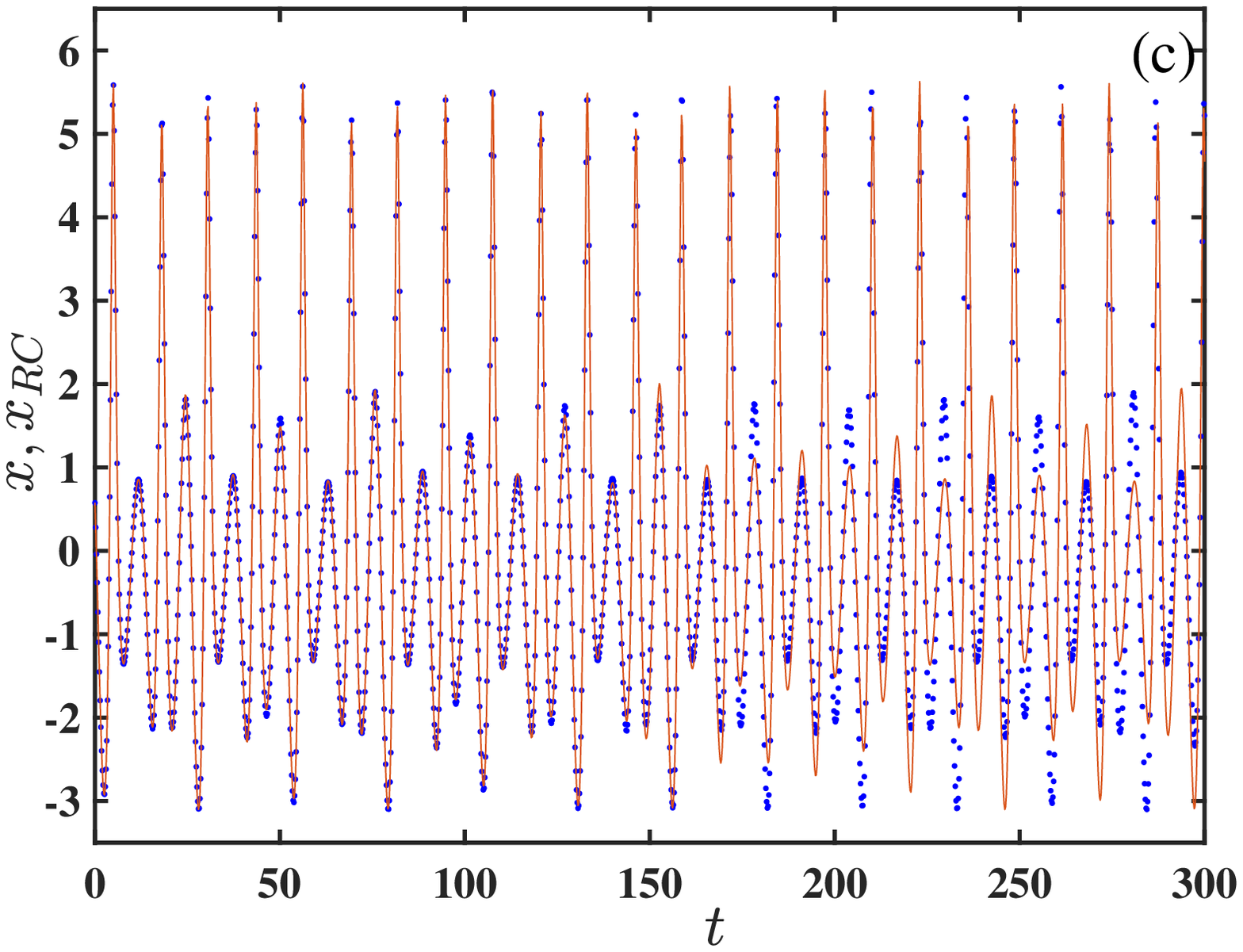}~\includegraphics[width=1.6in]{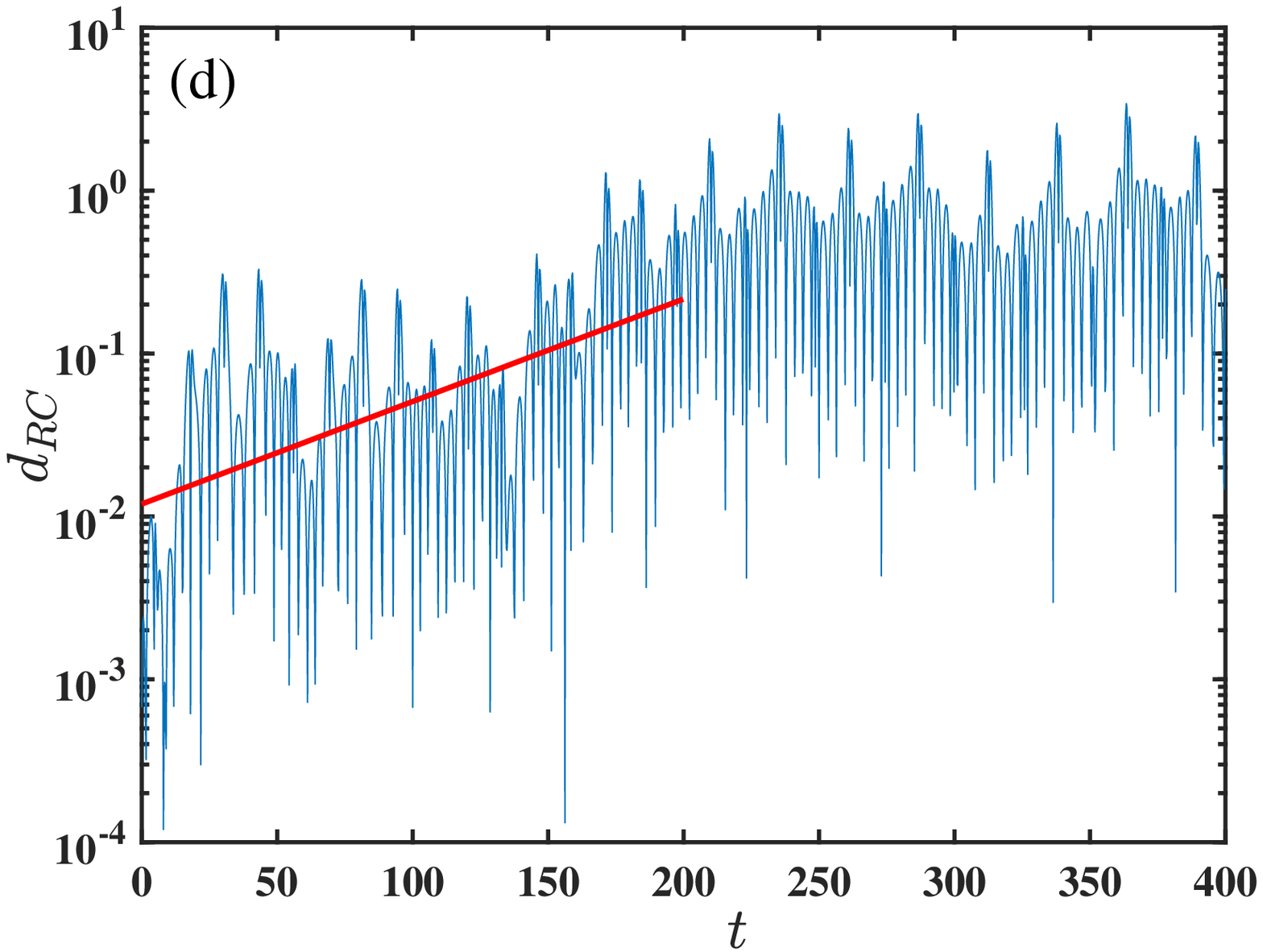}
	\caption{\label{fig4} The ESN trajectory (blue dotted line) is shown alongside a reference trajectory (red solid line) for Lorenz (a) and GIN (c) systems. In figures (b) and (d), the prediction error $d_{RC}(t)$ is shown for Lorenz and GIN systems, respectively. In these figures, the ordinate axis is shown in the logarithmic scale, and a straight fitting line is shown for checking the exponential scaling of the error. }
\end{figure}

The predictability of chaotic dynamics using an RC model was reported in many publications\cite{Jaeger:04,Pathak:18,Zimmermann:18,Fan:20, Weng:19,Kong:21,Kobayashi:21}. For example, Jaeger et al\cite{Jaeger:04} reported a very long prediction horizon $\tau=600$ for Lorenz system. Note that the same values of Lorenz system's parameters but different initial conditions are used in this work. The Jaeger et al\cite{Jaeger:04} considered a single trajectory obtained by 5th order Runge-Kutta scheme without any stochastic perturbations. Their RC model has been extensively optimised. In general, the development and implementation of an RC model involve a try and error approach\cite{Jaeger:04,Lukosevicius:12}. A predicted trajectory, $x_{RC}(t)$, for Lorenz system using ESN (\ref{eq3}-\ref{eq4}) is shown by dots in Fig.~\ref{fig4}(a). A reference trajectory used for the training and prediction is the same as that obtained by the classical 4th order Runge-Kutta scheme (blue dotted line in Fig.~\ref{fig1}(a)). The prediction horizon is $\tau\approx 5$ which is less than that for trajectories in Fig.~\ref{fig1}(a) and Fig.\ref{fig2}(a). The applied ESN consists of 500 units. The trajectory of  prediction error $d_{RC}(t)=|x(t)-x_{RC}(t)|$ (Fig.~\ref{fig4}(b)) includes an initial exponential growth with the factor $0.8$. This value is close to the largest Lyapunov exponent. The error $d_{RC} (t)$ reaches the size of the attractor after the prediction horizon. It means that the prediction after this interval is not possible. The application of a similar ESN for GIN system is presented in Fig.~\ref{fig4}(c) and (d). After $\tau\approx 171$, the prediction error $d_{RC}(t)$ reaches a saturation level. However, similarly to the numerical and stochastic noise cases, the error $d_{RC}(t)$ remains finite and oscillates along the reference trajectory. The prediction horizon and dynamics of the error $d_{RC}(t)$ are close to those for the stochastic approach. Note that the initial growth interval in $d_{RC}(t)$ is different from a single exponent.

\begin{figure}
	\noindent \includegraphics[width=1.6in]{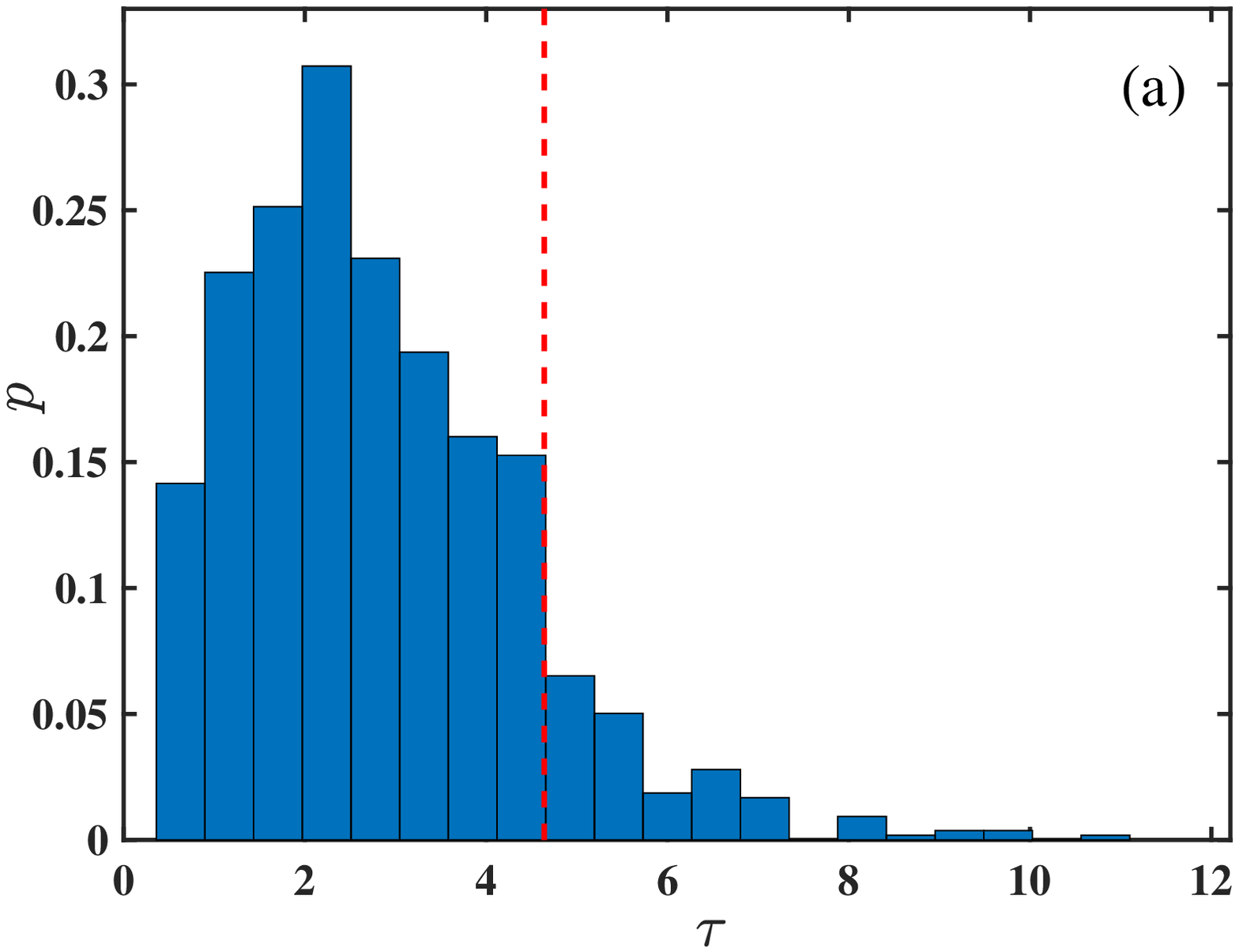}~~~\includegraphics[width=1.6in]{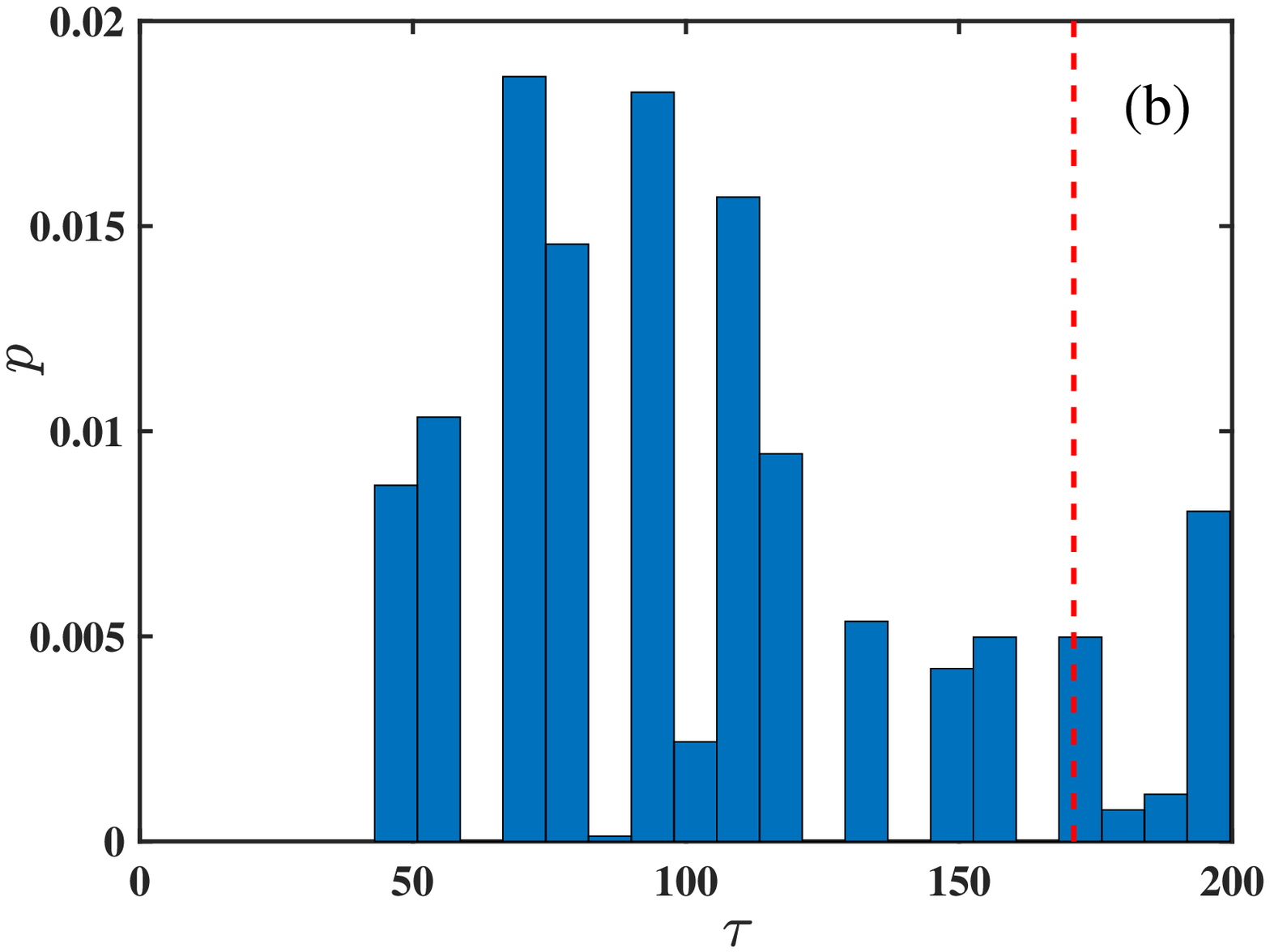}
	\caption{\label{fig5} The probability density, $p(\tau)$ of the RC prediction horizon for ensemble of stochastic trajectories of Lorenz (a) and GIN (b) systems. A vertical red dashed line corresponds to the prediction horizon obtained for deterministic trajectories shown in Figs.~\ref{fig4}(a) and (c). An ensemble of 1000 stochastic trajectories started with the same initial conditions was used.}  
\end{figure}

The results shown in Fig.~\ref{fig4} were obtained for a single trajectory that includes a numerical noise. For eliminating this noise and making trajectory more realistic, the stochastic approach is extended for the RC model. The ESN prediction was tested using an ensemble of stochastic trajectories started with the same initial conditions as those in Fig.~\ref{fig4}. The probability densities of the prediction horizon $\tau$ for both systems are shown in Fig.~\ref{fig5}. The stochastic perturbations lead to spreading $\tau$ in a wide range. On average, the horizon reduced for stochastic trajectories in comparison with the noise-free case (dashed vertical lines in Fig.~\ref{fig5}(a, b)), but some trajectories show a more extended prediction. Note that the maximal $\tau$ is less than the horizon obtained by the stochastic approach for both systems. This result is consistent with the fact that the stochastic approach provides an upper boundary on the maximal prediction horizon. Figure \ref{fig5} confirms the importance of the inclusion of the stochastic perturbations into the prediction estimation.

\section{\label{sec4}  Surrogate modelling using an RC model}

In recent publications\cite{Weng:19,Kong:21,Kobayashi:21}, an RC model was applied for surrogate modelling, which is widely used in engineering for replacing computationally expensive simulations\cite{Forrester:08}. The applicability of such modelling depends on how well a machine learning model generalises the knowledge extracted from data. Additionally, there is an implicit assumption that the system has a unique, robust state. Typically, surrogate modelling used for spatial systems with an equilibrium state. Therefore, the application of surrogate modelling to a chaotic system leads to new challenges. The main question is what the necessary criteria for a robust surrogate model are. For example, Kobayshi et al\cite{Kobayashi:21} showed that an RC model of Lorenz system preserves many properties of chaotic attractors. A possibility for predicting some bifurcations by an RC model was illustrated in work of Kong et al\cite{Kong:21}.

\begin{figure}
	\noindent \includegraphics[width=1.6in]{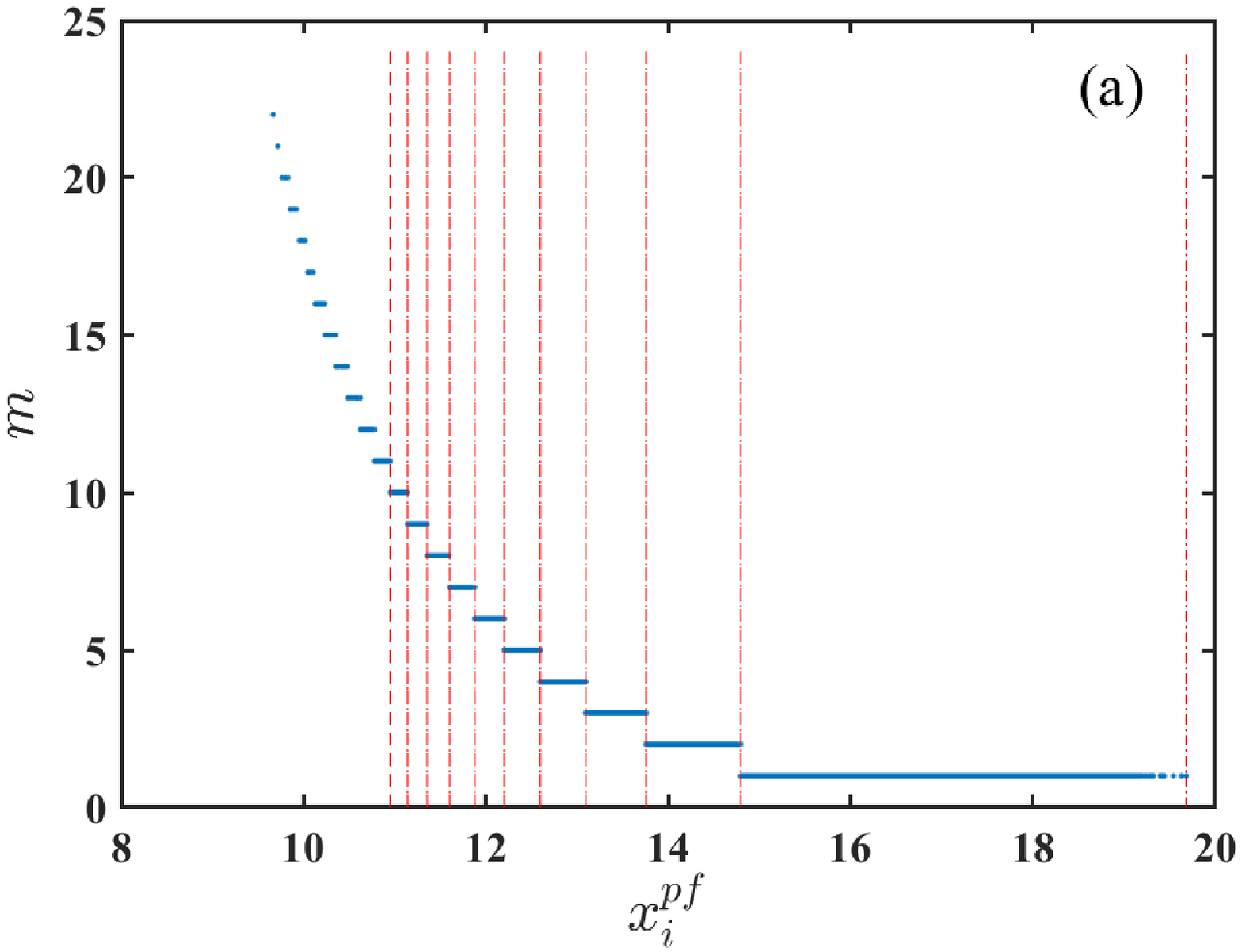}~\includegraphics[width=1.6in]{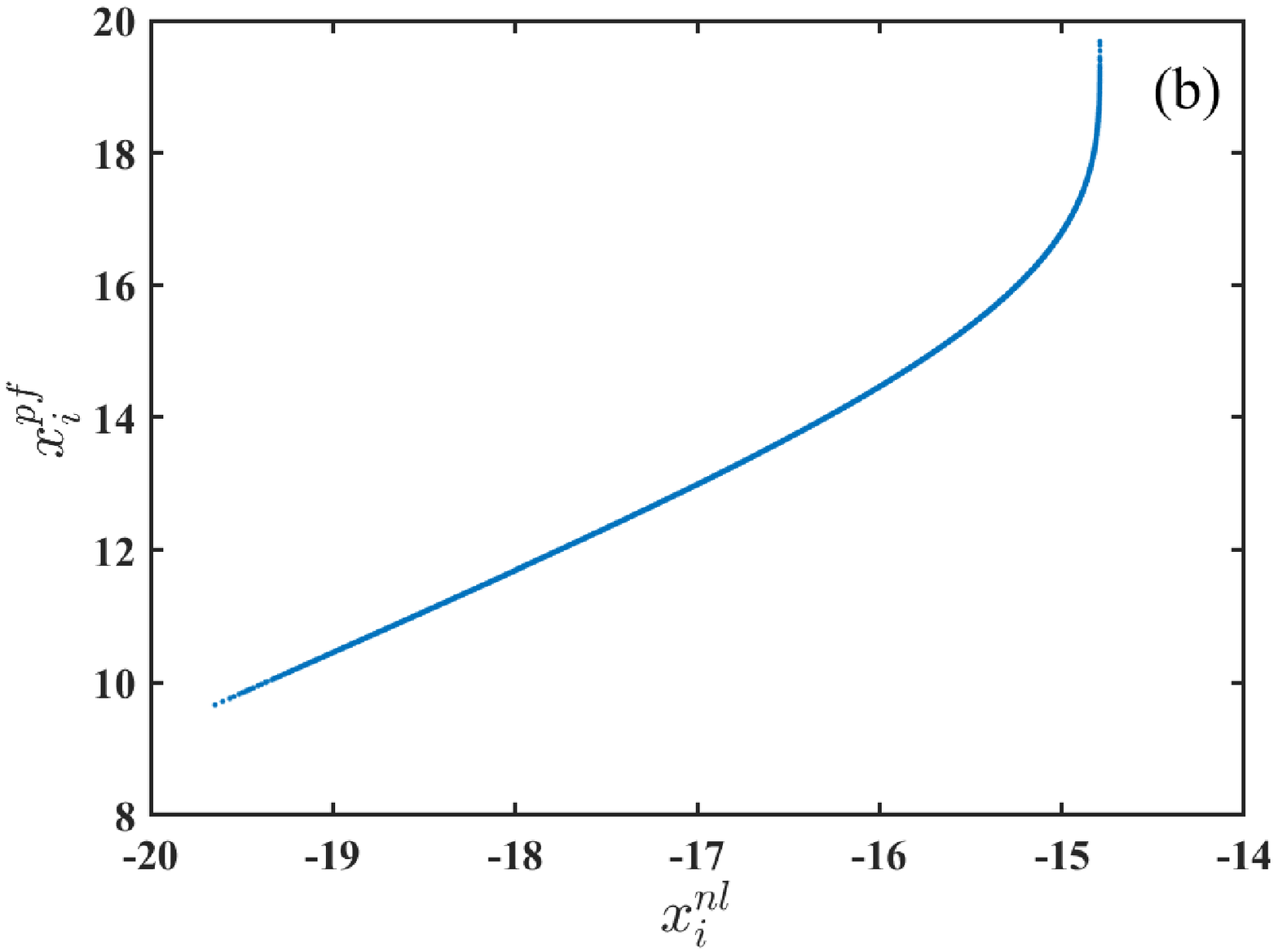}\\
	\noindent \includegraphics[width=1.6in]{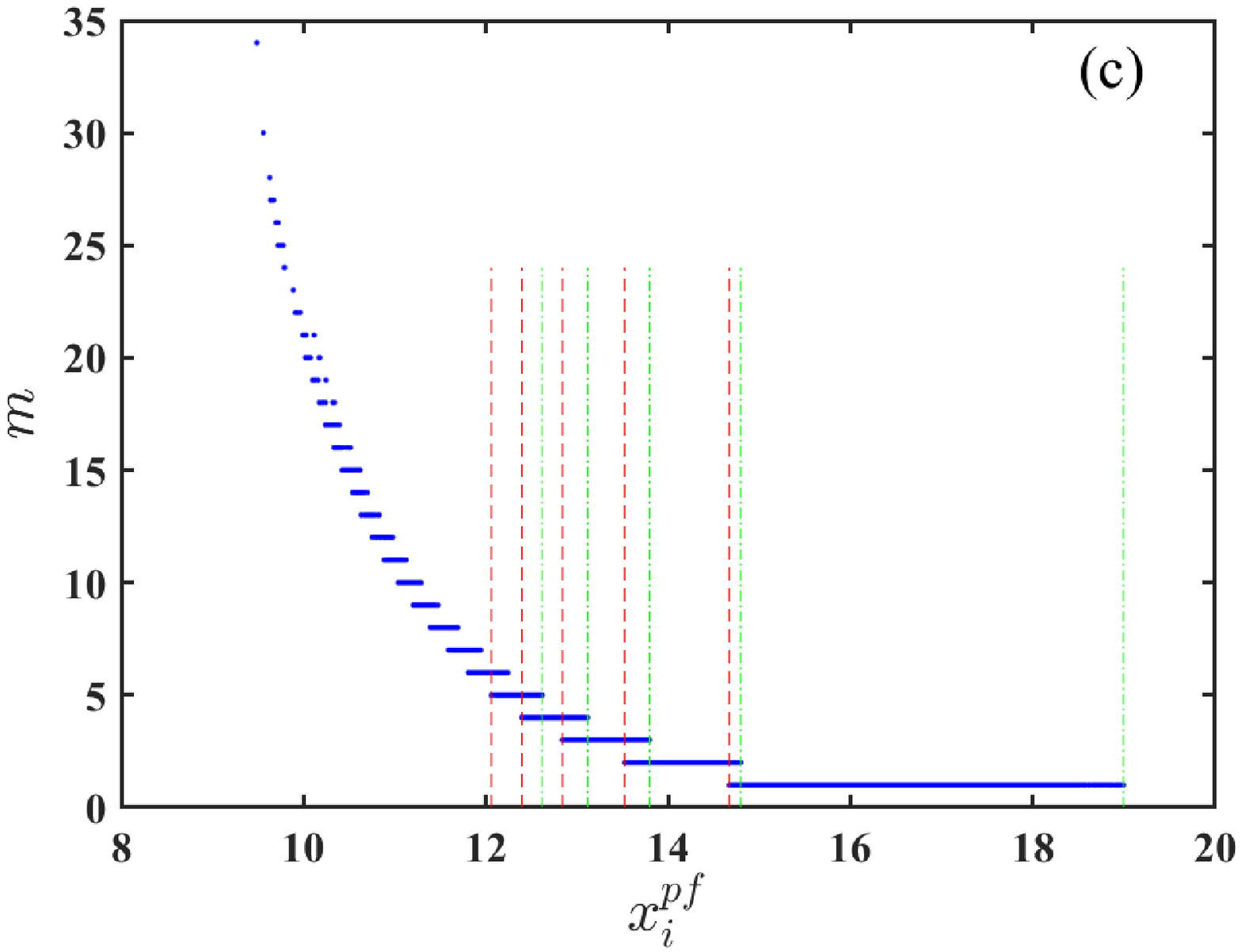}~\includegraphics[width=1.6in]{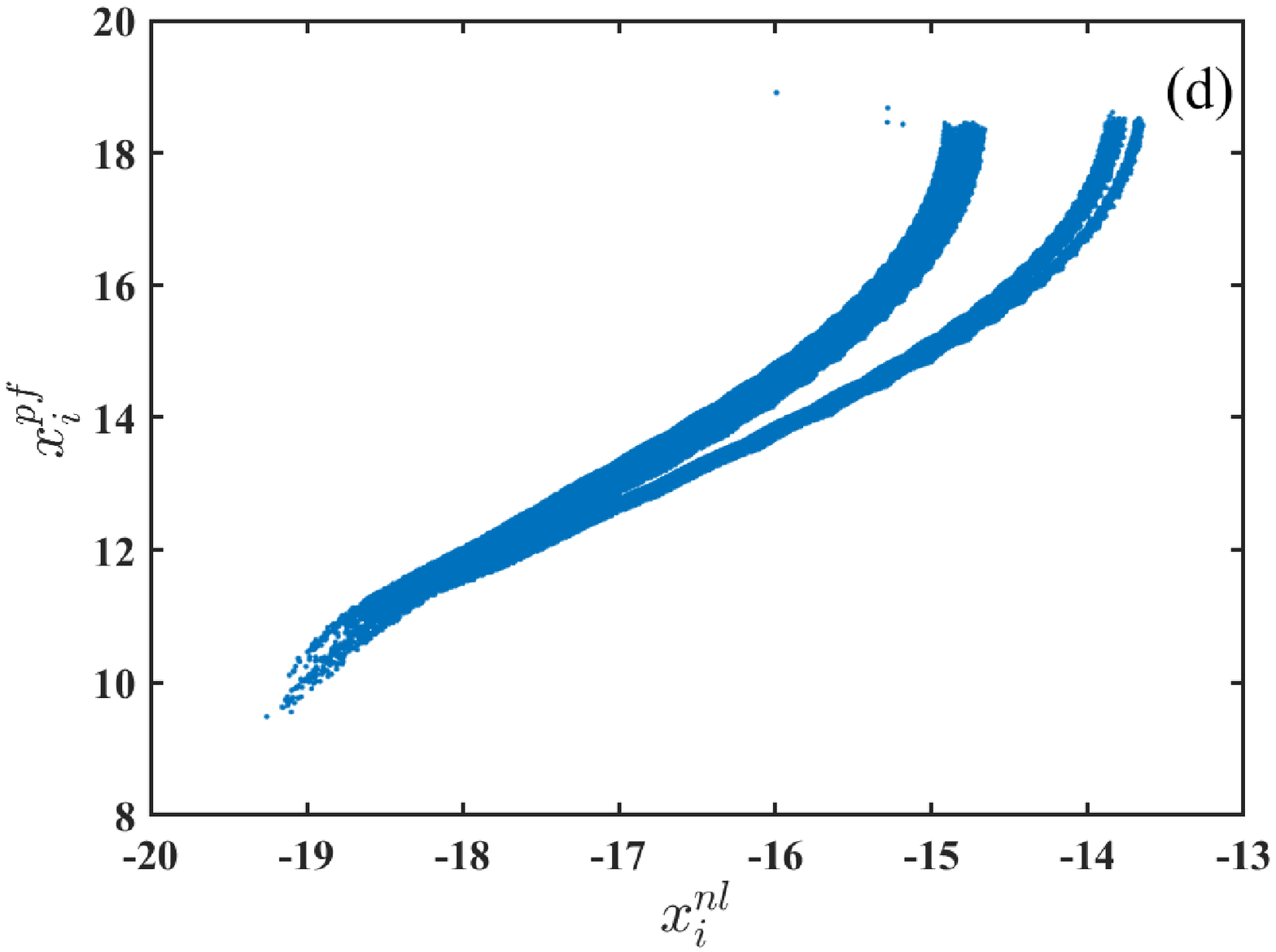}
	\caption{\label{fig6} The features of  Lorenz attractor are obtained using time series of  Lorenz system (a, b) and the ESN (c, d). The dependences of rotation number, $m$, on the value of the first extrema $x^{pf}_i$ are shown in figures (a) and (c). Red dashed and green dash-dotted vertical lines mark left and right boundaries of intervals, respectively. The relationship between values of $x^{nl}_i$ and $x^{pf}_i$  are shown in figures (b) and (d). }
\end{figure}

In this paper, parameters of Lorenz system are selected in the region of  Lorenz attractor existence. The attractor is one attracting state in the state space, and its structure is robust to small parameters' variations and stochastic perturbations. Therefore,  Lorenz attractor is a good candidate for surrogate modelling. The ESN was developed using a single time series of coordinate $x(t)$ of (\ref{eq5}). Several unique features of  Lorenz attractor can be derived based on time trace $x(t)$. For example, $x(t)$ can be converted to a sequence $x_i$ of extrema $|x(t)|$. One part $x^p_i$ of this sequence corresponds to positive and another part $x^n_i$ to negative values of $x(t)$. When trajectory $x(t)$ change the sign, it makes several rotations, $m$, (see Fig.~\ref{fig1}(a)) for an example).  Lorenz attractor structure defines a step-wise dependence (Fig.~\ref{fig6}(a)) of rotations number, $m$, on values of first extrema, $x^{pf}_i$, (or $x^{nf}_i$) observed after trajectory, $x(t)$, changes the sign. There are non-overlapping regions of $x^{pf}_i$ for each value of $m$. It means that the number of rotations, $m$, and, therefore, the dynamics can be predicted using the knowledge of the attractor structure. The boundary between regions is formed by the stable manifolds of the saddle point\cite{Jackson:90}. Another feature is nearly 1-dimensional monotonous dependence (Fig.~\ref{fig6}(b)) between values of $x^{pf}_i$ and $x^{nl}_i$, where the latter is the last extrema before trajectory changes the sign from negative to positive. Both features confirm that a trajectory of  Lorenz attractor can be predicted using a simple time series analysis. The application of the same analysis to the ESN trained time series of $x(t)$ (Fig.~\ref{fig4}(a)) is illustrated in Figs.~\ref{fig6} (c) and (d). It is apparent that the structure of  Lorenz attractor is not completely preserved in the ESN. Thus, the dynamics of the RC model is distinct. Therefore, an RC model for surrogate modelling should be applied with extreme caution. Note that differences between  Lorenz system and a trained ESN were observed in the study of synchronisation of the ESN driven by  Lorenz system\cite{Weng:19}. Weng et al \cite{Weng:19} showed that the synchronisation is observed even for a parameter mismatch. However, the analysis of saddle cycles of chaotic attractor in  coupled Lorenz systems\cite{Anishchenko:98} led to the conclusion that the complete synchronisation is not observed if systems parameters are different.

\begin{figure}
	\noindent \includegraphics[width=3in]{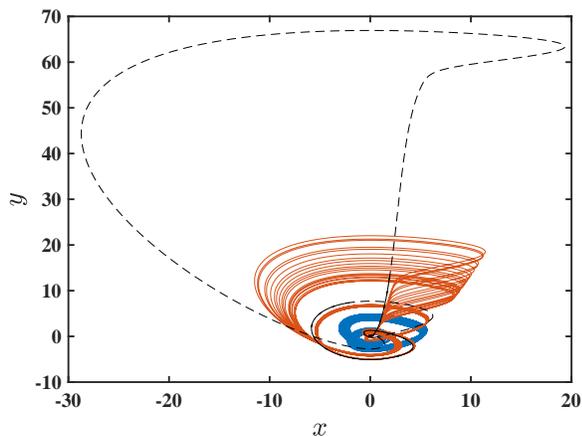}
	\caption{\label{fig7} The attractors of GIN system are shown on the state plane $x$-$y$. Two chaotic attractor are shown by blue and orange solid lines. The cycle of period 6 is shown by black dashed line. Low amplitude chaotic attractor (blue solid line) is used for developing an RC model. }
\end{figure}

The assessment of the surrogate modelling for GIN system is more problematic. The key feature of systems with the quasi-attractor is multistability. That is, different attractors can be observed by varying initial conditions. For selected parameters values, the system has three different attractors: two chaotic attractors and one cycle of period 6 (Fig.~\ref{fig7}). The ESN network was trained using a trajectory of one chaotic attractor only. The RC model was not able to reproduce other attractors and showed instability if the initial conditions are changed. Note that the noise intensities used in the stochastic approach were small, and the stochastic perturbations did not induce transitions between the attractors.

\section{\label{sec5}  Conclusion }

The results stress the importance of selecting an appropriate framework for considering the predictability of chaotic time series. A numerical solution of ordinary differential equations includes an embedded complicated multiplicative noise that depends on a particular numerical scheme. A high-order scheme reduces the numerical error but does not eliminate it. Low-order schemes such as the Euler method have a larger error which significantly affects the dynamics. Note that reducing the step size in the Euler method for achieving a comparable accuracy with high-order schemes, for example, the classical 4th order Runge-Kutta scheme, is often computational prohibitive because of the power-law dependence of the error on the step size. As a result, a numerical time series represents both the system and a numerical scheme. For addressing this issue, a stochastic approach with explicit additive noise has been suggested. The approach effectively switches the consideration to SDEs. In the approach, the intensities of the stochastic perturbations are selected to be comparable with the numerical noise. The latter is linked to the step size of a numerical scheme. Also, the stochastic approach requires a statistical consideration using an ensemble of stochastic trajectories. In such a case, the predictability of a chaotic system started with particular initial conditions is characterised by the ensemble's time evolution. The range of ensemble trajectories gives the predictability horizon, which provides an upper boundary on a meaningful prediction interval. Real systems are stochastic, and the approach includes this property explicitly. The presented results show a distinct predictability pattern in Lorenz and GIN systems. A small error induced by either numerical or stochastic noise reaches the attractor size in Lorenz system. However, in GIN system, the error remains finite for a very long time interval. These patterns reflect the structure of these systems' state spaces.

An RC model in the form of ESN (\ref{eq3}) shows similar predictability patterns as those given by the stochastic approach. In the presence of stochastic perturbations, the ESN prediction horizon is varied in a wide range for both systems (Fig.~\ref{fig5}). So, ESN training shows a sensitivity to the noise presence. If a solution of a deterministic system is used for ESN training, then the RC model reflects the noise of a numerical scheme. It means that ESN's prediction horizon depends on the numerical scheme used. Note that real chaotic systems are inherently stochastic. For example, GIN system describes an electronic circuit with different noise sources from circuit's components. Therefore, SDEs and the corresponding statistical description should be used for developing and testing a predictive model.
 
A brief assessment of the generalisation and possibility of surrogate modelling using RC highlighted significant RC's misrepresentation of the system state space. This observation is not a surprise since a relatively short time series was used on the training stage. The size of training data is small to avoid overtraining of the ESN. Different techniques are required for developing an RC model for surrogate modelling. Also, there is no good understanding of ESN properties and the influence of the network's parameters. The selection of suitable training parameters is a time-consuming process, which does not guarantee an optimal choice. Since an RC model represents an analogue computer, the fluctuations should be included in the model. 

In the paper, fluctuations acting on chaotic systems are small, i.e. of the order of numerical noise. Real systems, however, are characterised by a considerably higher level of noise. In this case, the noise influence can be strong and can dramatically change the properties of chaotic systems. Developing an RC model for chaotic systems with significant noise could be a challenge since a critical question of what is chaos in the presence of noise remains outstanding. Therefore, the answer to this question and the characterisation of the RC predictability and interpretation of RC modelling of chaotic systems with strong noise require further research.

\begin{acknowledgments}
	The author dedicates this work to the memory of his teacher, Vadim Semenovich Anishchenko.
\end{acknowledgments}

\section*{Data Availability Statement}

The data that support the findings of this study are openly available in wrap.warwick.ac.uk, reference number 153069 with the URL http://wrap.warwick.ac.uk/153069/ .

\nocite{*}
%

\end{document}